\documentclass[aps,pre,preprint,showpacs,showkeys,12pt]{revtex4-2}
\usepackage{bm}
\usepackage{graphicx}
\usepackage{float}
\usepackage{caption}
\usepackage{subcaption}
\usepackage{amsmath}
\usepackage{amssymb}
\usepackage{comment}
\usepackage{appendix}

\begin{document}


\title{Theoretical Study of Inhomogeneity Effects on Three-Wave Parametric Instability: A WKBJ Approach } 



\author{Taotao Zhou}
    \email{zhoutaotao@ipp.ac.cn}
    \affiliation{Institute of Plasma Physics, Hefei Institutes of Physical Science, Chinese Academy of Sciences, Hefei 230031, China} 
    \affiliation{University of Science and Technology of China, Hefei 230026, China}
\author{Nong Xiang}
   \affiliation{Institute of Plasma Physics, Hefei Institutes of Physical Science, Chinese Academy of Sciences, Hefei 230031, China} 
\author{Chunyun Gan}
    \affiliation{Institute of Plasma Physics, Hefei Institutes of Physical Science, Chinese Academy of Sciences, Hefei 230031, China} 
\author{Tianyang Xia}
    \affiliation{Institute of Plasma Physics, Hefei Institutes of Physical Science, Chinese Academy of Sciences, Hefei 230031, China} 




\date{\today}

\begin{abstract}
ABSTRACT.
The mechanisms by which media inhomogeneity affects the three wave parametric instability (PI), including the wave number mismatch and the parameter gradients, are investigated using an approach based on the Wentzel-Kramers-Brillouin-Jeffreys (WKBJ) approximation. This approach transforms the coupling wave equations into an amplitude equation and iteratively solves its characteristic polynomials.
By analyzing the solutions, we proposed that the wave number of the quasi-mode, a key term in the wave number mismatch of non-resonant type PI, should be a complex root of the quasi-mode's linear dispersion equation. Based on this, we derive a unified amplification factor formula that covers the resonant and non-resonant, the forward-scattered and backward-scattered types of PI. The impact of parameter gradients on the local spatial growth rate becomes significant when the inhomogeneity exceeds $10^{-3}$. Considering parameter gradients extends our approach's validity to an inhomogeneity of about $10^{-2}$. 
This approach holds promise for more specific PI modeling in the future.
\end{abstract}


\maketitle 

\section{Introduction}
Parametric instability (PI) is a ubiquitous nonlinear phenomenon extensively observed and studied across various fields, including fluid dynamics\cite{edwardsPhys.Rev.E1993}, nonlinear acoustics\cite{maccariJournalofSoundandVibration2001,versluisPhys.Rev.E2010}, nonlinear optics\cite{trilloPhys.Rev.E1997}, and plasma physics\cite{leePhys.Rev.E1999}. Typically, PI involves the excitation of a mode with an intrinsic frequency of the system by external modes at different frequencies through nonlinear mode coupling. When the system is a medium that non-linearly responds to perturbations, PI can manifest as three-wave interactions. A three-wave PI refers to the process where a pump wave with frequency and wave number $(\omega_0,k_0)$ decays into a child mode 'u' ('u' means undetermined) with $(\omega_u,k_u)$ and a resonant child wave '1' with $(\omega_1,k_1)$. The frequencies and wave numbers of these three waves satisfy the matching conditions $\omega_1=\omega_u\pm \omega_0$ and $k_1=k_u\pm k_0$. \cite{kaupStudApplMath1976}

There are several ways to classify different types of PI, depending on the characteristics of the child waves. Resonant type PI is characterized by both excited child waves being resonant waves \cite{stenfloPhys.Scr.1994} , whereas non-resonant type\cite{liuPhys.Rep.-Rev.Sect.Phys.Lett.1985} indicates that the child wave 'u' is a quasi-mode with strong damping. Non-resonant type PI is commonly observed in plasma environments due to its ability to sustain intense damping modes, such as electron Landau damping and ion cyclotron damping\cite{swanson2003}. Forward-scattered PI refers to the scenario where child waves propagate in the same direction, meaning the group velocities of child waves satisfy $v_{gu}v_{g1}>0$. Conversely, backward-scattered PI means the child waves propagate in an opposite direction ($v_{gu}v_{g1}<0$).\cite{porkolabPhys.Fluids1970}  

The three-wave PI has a rich history in theoretical studies. In the beginning, researchers emphasized PI in homogeneous media.\cite{kaupRev.Mod.Phys.1979} The governing equations for the three-wave PI are the coupling wave equations. In a homogeneous medium where parameters and matching conditions remain constant, these can be transformed into the parametric dispersion relations, expressed as $\varepsilon_1\varepsilon_u-\alpha\alpha^*=0$. Here, $\varepsilon_1$ and $\varepsilon_u$ represent the linear dispersion relations of child waves '1' and 'u', respectively, while $\alpha$ and $\alpha^*$ are the coupling coefficients between the pump wave and the child waves '1' and 'u'. By performing a first-order expansion of the linear dispersion relations of the resonant child waves, such as $\varepsilon_u\simeq i\gamma\frac{\partial \varepsilon_u}{\partial \omega_u}$, one can derive the growth rate\cite{liuPhys.Rep.-Rev.Sect.Phys.Lett.1985} 
\begin{equation}
    \gamma\left.\right|_{\mathrm{non-resonant}}=\frac{\mathrm{Im}\left[\alpha\alpha^*/\varepsilon_u(k_u)\right]}{\partial \varepsilon_1/\partial \omega_1}, 
    \label{Eq_GamNr}
\end{equation}
\begin{equation}
    \gamma_0^2\left.\right|_{\mathrm{resonant}}=\frac{-\alpha\alpha^*}{\frac{\partial \varepsilon_u}{\partial \omega_u}\frac{\partial \varepsilon_1}{\partial \omega_1}}
    \label{Eq_GamRe}
\end{equation}
for non-resonant and resonant types of PI, respectively. The growth rate is an important indicator of the local strength of PI. 

However, in a non-uniform environment, various new factors come into play that influence PI amplification. This makes it challenging to rely solely on the growth rate to predict PI behavior accurately. As the excited child wave propagates through the inhomogeneous medium, local parameters will vary, altering the intrinsic wave number, linear propagation characteristics, and PI coupling strength. Changes in wave numbers can disrupt the matching conditions, thereby inhibiting PI growth. Variations in group velocity also affect the amplitude and phase of the child wave. These factors must be comprehensively studied to understand PIs in inhomogeneous media.

In the early 1970s, Rosenbluth established a pioneering theory describing PI under inhomogeneous conditions for the resonant type,\cite{rosenbluthPhys.Rev.Lett.1972,rosenbluthPhys.Rev.Lett.1973} where the inhomogeneity is introduced by expanding the linear dispersion relation near the matching point, represented as $\varepsilon_1\Phi_1 \to \left(\varepsilon_1-i \frac{\partial \varepsilon_1}{\partial k_1}\frac{\partial}{\partial x}\right) \Phi_1$. This approach highlights the inhibitory effect of the wave number mismatch $\kappa$ on PI growth. It is found that child waves can only grow within a limited region near the matching point, where $\kappa$ is below a critical threshold. This spatial restriction on PI coupling imposes an upper limit on the amplification of the child waves, commonly referred to as the amplification factor
\begin{equation}
    A_{\mathrm{inhom}}=\dfrac{\pi\gamma_0^2}{\frac{\partial \varepsilon_u}{\partial \omega_u}\frac{\partial \varepsilon_1}{\partial \omega_1}\kappa^\prime}.
    \label{Eq_AinhomRe}
\end{equation}
This upper limit also serves as a crucial indicator for estimating the strength and threshold of PI growth in inhomogeneous media. 

For non-resonant type PI under inhomogeneity, it is proposed that there should be an amplification factor similar to that of the resonant type, though its growth is not strictly confined near the matching point as in the resonant case. This amplification factor has been derived using various methods\cite{liuPhys.Rep.-Rev.Sect.Phys.Lett.1985,cesarioNucl.Fusion2006,zhaoPhys.Plasmas2013} , resulting in the established expression
\begin{equation}
    A_{\mathrm{inhom}}=\dfrac{\pi\gamma\mathrm{Im}(\varepsilon_u)}{v_{g1}\kappa^\prime}\left[\mathrm{Re}\left(\frac{\partial \varepsilon_u}{\partial k_u}\right)\right]^{-1}.
    \label{Eq_AinhomNr}
\end{equation}
However, this expression has some unclear aspects. The child mode 'u' is a quasi-mode, meaning its wave number $k_u$, which is also an essential parameter of $A_{\mathrm{inhom}}$, cannot be accurately determined and is arbitrarily chosen since $\varepsilon_u(k_u)\neq 0$. About the choice of $k_u$, one hypothesis is that $k_u$ should meet the matching condition $k_u=\pm k_0+k_1$ wherever its relevant PI is triggered, implying that wave number mismatch $\kappa$ is constantly zero and thus does not affect PI growth.\cite{takasePhys.Fluids1983} This hypothesis offers a fundamentally different physical interpretation of the non-resonant type PI compared to the resonant type, making it challenging to explain the physics of an intermediate type between them. Another hypothesis suggests choosing a $k_u$ that makes the child wave '1' propagate as similar as possible to the pump, thereby minimizing the impact of wave number mismatch.\cite{cesarioNucl.Fusion2006} Later in this work, we will also conduct a comprehensive analysis on this issue, where we will not presume a specific value for $k_u$ but derive it from first principles. 

Currently, precise PI modeling becomes essential for the large-scale integration simulations and engineering control.\cite{cesarioPhys.Rev.Lett.2004,bao2020} Accurate prediction of child wave amplification within specific intervals of its trajectory under various practical circumstances is needed.
A direct approach\cite{cesarioNucl.Fusion1989} is to integrate over the concerned region. Consider an interval $[x_a,x_b]$ for instance, the growth of child wave is described by $\Phi(x_b)=\Phi(x_a) \exp\left[ \int_{x_a}^{x_b} \frac{\gamma}{v_{gx}}dx \right]$, where $\gamma/v_{gx}$ represents the spatial growth rate in the $x$ direction. However, this method does not account for the effect of parameter gradients and lacks a basis in first principles. The spatial growth rate will also increase to an unrealistic infinity when the child wave encounters a mode conversion or reflection layer where $v_{gx}\to 0$.
A better approach involves directly using numerical computations to solve the coupling wave equations or those simplified by partial analytical approximations.\cite{napoliJ.Phys.Conf.Ser.2012,napoliPlasmaPhys.Control.Fusion2013} Sometimes, these methods might not be universally applicable as they depend on approximations designed for particular scenarios and boundary conditions.  

Drawing from the historical and current understanding of PI theory in inhomogeneous media, there remains a need for deeper insight into the physical image and mathematical modeling. To comprehensively include the effects induced by inhomogeneity, this work delves into studying PI based on first principles, i.e., the coupling wave equation. The WKBJ approximation is a valuable mathematical technique for simplifying the wave equations. It involves assuming the medium parameters and the wave amplitude varies much more slowly than the eikonal, and has been used to address wave equations focusing on a single mode.\cite{swanson2003} In this work, we extend the WKBJ approximation to solving the coupling wave equations. This approximation will be crucial for separating the amplitude and eikonal of child wave '1', as well as solving the amplitude equation. 

The structure of this paper is outlined as follows: Sec. \ref{sec_2} presents the derivation to solve the coupling wave equations, establishing the relationship between the amplitude equation and the parametric dispersion relation. Sec. \ref{sec_3} explores the mechanism of wave number mismatch, focusing on the unclear quasi-mode's wave number by recalculating the amplification factors of different PI types. Sec. \ref{sec_4} examines the effect of the parameter gradients on the local spatial growth rate of PI. Numerical examples are included at the end of Sec. \ref{sec_3} and Sec. \ref{sec_4} to verify the growth behavior of child wave '1' in different types of PI and to confirm the applicability range of our method regarding inhomogeneity, respectively. Sec. \ref{sec_5} summarizes the main steps of our approach and the revised physical insights derived from its solutions.

\section{Approach for solving coupling wave equations\label{sec_2}}


\subsection{Derivation of the amplitude equation}

In an inhomogeneous 1D medium, consider the PI process involving three monochromatic waves $(\omega_0,k_0)\Rightarrow (\omega_1,k_1) + (\omega_u,k_u)$. The governing coupling equations with finite highest derivative order can be written as 
\begin{equation}
    \sum_nf_n\phi_0^{(n)}=\sum_{m_1,m_u} a_{m_1,m_u}\phi_u^{(m_u)}\phi_1^{*(m_1)},
\end{equation}
\begin{equation}
    \sum_n b_n \phi_u^{(n)}=\sum_{m_0,m_1} a_{m_0,m_1} \phi_0^{(m_0)} \phi_1^{(m_1)},
\end{equation}
\begin{equation}
    \sum_{n} c_n\phi_1^{(n)}=\sum_{m_0,m_u} a_{m_0,m_u} \phi_0^{*(m_0)}\phi_u^{(m_u)}.
\end{equation}
where $\phi$ denote the perturbation of each wave. The superscript $(n)$ with brackets denotes the $n$th spatial derivative $\frac{d^n}{dx^n}$. Coefficients $f_n$, $b_n$ and $c_n$ represent the medium's linear response to each wave, while tensor $\boldsymbol{a}$ on the right-hand side represents the medium's quasi-linear or nonlinear response. \cite{benneyJournalofMathematicsandPhysics1967} For simplicity, we assumed a stationary condition and all time derivatives $\frac{\partial}{\partial t}$ act only on the time-dependent term $\exp(-i\omega t)$. The produced  $-i\omega$ terms have been included in coefficients of the above equations.


Suppose that the amplitude of the pump wave $\Phi_0$. is sufficiently large and its eikonal is primarily determined by the known wave number $k_0(x)$. This ensures that the pump remains nearly unaffected by the coupling. Under these conditions, the three-wave equations are simplified to a form of two-wave coupling
\begin{equation}
    \sum_{n=0}^N b_n \phi_u^{(n)}=\sum_{m=0}^M a_m e^{i\Psi_0} \phi_1^{(m)},
\label{Eq_CoupWave1}
\end{equation}
\begin{equation}
    \sum_{n=0}^{N_1} c_n\phi_1^{(n)}=\sum_{m=0}^M a_m^* e^{-i\Psi_0} \phi_u^{(m)},
    \label{Eq_CoupWave2}
\end{equation}
where the pump's amplitude $\Phi_0$ has been included in the coupling parameters $a_m$ and $a_m^*$. The highest derivative order of the equations may vary with specific physical conditions. For example, in simple parametric oscillators, $\text{N}=\text{N}_1=2$, while $\text{M}$ is determined by the pumping force. In some isotropic media, such as magnetized plasma, $\text{N}$ and $\text{N}_1$ may be larger than 2 for certain cyclotron waves, and $\text{M}$ may exceed 1 for coupling perpendicular to the magnetic field. 

To solve Eqs. (\ref{Eq_CoupWave1},  \ref{Eq_CoupWave2}), the key step involves eliminating the unknown variable $\phi_u$, whose intrinsic wave number $k_u$ and eikonal $\Psi_u$ are not explicitly known. The elimination is achieved by differentiating Eq. (\ref{Eq_CoupWave1}) $\text{M}$ times and Eq. (\ref{Eq_CoupWave2}) $\text{N}$ times. The resulting set of linear equations is expressed as 
\begin{equation}
    \left(\begin{array}{cc}
\boldsymbol{M}_{b} & \boldsymbol{V}_{1a}\\
\boldsymbol{M}_{a^{*}} & \boldsymbol{V}_{1c}
\end{array}\right)\left(\begin{array}{c}
\phi_{u}\\
\phi_{u}^{\prime}\\
\vdots\\
\phi_{u}^{(M+N)}\\
-1
\end{array}\right)=0.
\label{Eq_LinMat}
\end{equation}
Note that the last element of the $\phi_u$-vector is -1, indicating it is a non-zero vector. For the equations to be solvable, the determinant of the big combined matrix must be 0. Consequently, we obtain an equation without $\phi_u$ terms as
\begin{equation}
    \mathrm{det}\left|\begin{array}{cc}
\boldsymbol{M}_{b} & \boldsymbol{V}_{1a}\\
\boldsymbol{M}_{a^{*}} & \boldsymbol{V}_{1c}
\end{array}\right|=0.
\label{Eq_det}
\end{equation}

Let us analyze the components of the big determinant. The matrix $\boldsymbol{M}_{b}$ represents a $(\text{M}+1)*(\text{M}+\text{N}+1)$ matrix constructed from $b_n$ and its derivatives. Similarly, the matrix $\boldsymbol{M}_{a^*}$ is a $(\text{N}+1)*(\text{M}+\text{N}+1)$ matrix constructed from $a_m^*$ and its derivatives. We can decompose $\boldsymbol{M}_{b}$ or $\boldsymbol{M}_{a^*}$ based on the derivative order of the coefficients $b_n$ or $a_m^*$, giving $\boldsymbol{M}_{b}=\sum_{j=0}^M \boldsymbol{M}_{bj}$, $\boldsymbol{M}_{a^*}=\sum_{j=0}^N \boldsymbol{M}_{a^*j}$, where each component is 

\begin{equation}
    \boldsymbol{M}_{bj}=\left(\begin{array}{ccccc}
\\
C_{j}^{j}b_{0}^{(j)} & \cdots & C_{j}^{j}b_{N}^{(j)}\\
 & \ddots & \ddots & \ddots\\
 &  & C_{j}^{M+1}b_{0}^{(j)} & \cdots & C_{j}^{M+1}b_{N}^{(j)}
\end{array}\right)\begin{array}{c}
\\
\leftarrow\text{the }j\text{th row}\\
\\
\\
\end{array}.
\label{Eq_Mbj}
\end{equation}
The matrix $\boldsymbol{M}_{a^*j}$ is obtained by replacing all the $b_n$ in $\boldsymbol{M}_{bj}$ with $a_m^*$.

On the other hand, $\boldsymbol{V}_{1a}$ and $\boldsymbol{V}_{1c}$ is respectively a M+1 and N+1 length vector containing $\phi_1$-relevant terms. The $j$th component of $\boldsymbol{V}_{1a}$ is
\begin{equation}
    V_{1aj}=\sum_{n=0}^M\sum_{m=0}^j C_m^j a_n^{(j-m)}\left(\phi_1^{(n)}e^{i\Psi_0}\right)^{(m)}.
    \label{Eq_V1aj}
\end{equation}
Now we can use the WKBJ approximation to separate the wave perturbation $\phi_1$ into amplitude $\Phi_1$ and eikonal $\Psi_1$, i.e.,%
\begin{equation}
    \phi_1(x)=\Phi_1(x)e^{i\Psi_1(x)},
\end{equation}
where the eikonal $\Psi_1$ is a known variable given by $\Psi_1^\prime=k_1$. The additional change of eikonal due to the coupling has been included in the amplitude $\Phi_1$, which is the unknown variable to be solved.   


Use this expansion and the chain rule of differentiation, the term $\left(\phi_1^{(n)}e^{i\Psi_0}\right)^{(m)}$ in the Eq. (\ref{Eq_V1aj}) can be expressed in terms of $\Phi_1$ as
 
\begin{equation}
    \left(\phi_1^{(n)}e^{i\Psi_0}\right)^{(m)}=\sum_{m_a=0}^m\sum_p^{n+m_a}C_{m_a}^mC_p^{n+m_a}\Phi_1^{(p)}\left(e^{i\Psi_1}\right)^{(n+m_a-p)}\left(e^{i\Psi_0}\right)^{(m-m_a)},
\end{equation}

where $C_{m_a}^{m}$ and $C_p^{n+m_a}$ denote the binomial coefficients. Expanding $\boldsymbol{V}_{1a}$  as $\sum_p \boldsymbol{V}_{1ap}\Phi_1^{(p)}$ , we have each component of $\boldsymbol{V}_{1ap}$ as
\begin{equation}
    V_{1ajp}=\sum_{n=0}^M\sum_{m=0}^j C_m^j a_n^{(j-m)}\sum_{m_a=0}^mC_{m_a}^mC_p^{n+m_a}\left(e^{i\Psi_1}\right)^{(n+m_a-p)}\left(e^{i\Psi_0}\right)^{(m-m_a)}.
    \label{Eq_V1ajp}
\end{equation}
The components of $\boldsymbol{V}_{1c}$ have the same form as $\boldsymbol{V}_{1a}$, except for replacing $a_n$ with $c_n$.

The derivatives of the eikonal terms in Eq. (\ref{Eq_V1ajp}) can be expressed in terms of the known quantities $k_1$ and $k_0$. Generally, an $S$th order derivative on $\exp(i\Psi)$ is in a form similar to the exponential Bell polynomials as%
\begin{equation}
    \left(e^{i\Psi}\right)^{(S)}=e^{i\Psi}\sum_{u_j}S!\prod_j\dfrac{[(ik)^{(u_j-1)}/u_j!]^{p_j}}{p_j!},
    \label{Eq_IntPart}
\end{equation}
where $\{u_j\}$ represents a possible integer partition of $S$, and $p_j$ is the number of occurrences of integer $u_j$ in the partition scheme. These quantities satisfy the relationship  $S=\sum_j p_ju_j$. The summation with respect to $u_j$ encompasses all feasible integer partitions. %

Since $\boldsymbol{V}_{1a}$ and $\boldsymbol{V}_{1c}$ are expressed in terms of $\Phi_1$, we can obtain the target amplitude equation by substituting Eq. (\ref{Eq_V1ajp}) into the determinant in Eq. (\ref{Eq_det}), 
\begin{equation}
    \sum_p H_p\Phi_1^{(p)}=0
    \label{Eq_AmpEq}
\end{equation}
where the coefficient $H_p$ is
\begin{equation}
    H_p=\mathrm{det}\left|\begin{array}{cc}
\boldsymbol{M}_{b} & \boldsymbol{V}_{1ap}\\
\boldsymbol{M}_{a^{*}} & \boldsymbol{V}_{1cp}
\end{array}\right|.
\label{Eq_HpDet}
\end{equation}
The components of the determinant are obtained using Eqs. (\ref{Eq_Mbj}) and (\ref{Eq_V1aj}).

\subsection{Solutions of amplitude equation}

Since the amplitude equation (\ref{Eq_AmpEq}) is a linear ordinary homogeneous differential equation, we propose an iterative way to solve it using a series of characteristic equations. Suppose the solution has an exponential form $\Phi_1(x)=\Phi_1(0)\exp\left(\int_0^x R_1(\lambda) d\lambda\right)$, where $R_1=r_1+\delta r_1+\delta^2 r_1+...$. The real part $\text{Re}(R_1)$ signifies the local spatial growth rate of the child wave '1', a key quantity for comprehending PI. The zeroth-order component $r_1$ is defined as the minimal root of the characteristic polynomial $\sum_p H_pr_1^p=0$. In light of knowing $r_1$, one can transform the amplitude equation into a new equation as 
\begin{equation}
    \sum_l\left[\sum_p C^p_l H_p (e^{\int_0^x r_1d\lambda})^{(p-l)}\right]\delta \Phi_1^{(l)}=0,
\end{equation}
where $\delta \Phi_1=\Phi_1e^{-\int_0^x r_1(\lambda) d\lambda}=\exp\left[\int_0^x (\delta r_1+\delta^2 r_1+...)d\lambda\right]$ represents the remaining part of the solution to be determined. The content within the square bracket is denoted as $\delta H_l$. Consequently, we again encounter a linear ordinary differential equation similar to Eq.(\ref{Eq_AmpEq}), which can be solved by the characteristic method. Its characteristic polynomial as $\sum_p \delta H_p \delta r^p=0$ and $\delta r_1$ is one of roots (usually the smallest one). After obtaining $\delta r_1$, the same method is used iteratively to determine $\delta^2 r_1$, $\delta^3 r_1$ and higher-order terms. Generally, the characteristic polynomial of $n$th iterative order is
\begin{equation}
    \sum_p\delta^nH_p \delta^n r^p=0,    
\end{equation}
 where the coefficients follow an iterative relation 
\begin{equation}
    \delta^{n+1}H_p=\sum_m C^m_p \delta^n H_m\frac{ \left(e^{\int_0^x \delta^n r_1d\lambda}\right)^{(m-p)}}{e^{\int_0^x \delta^n r_1d\lambda}}.
    \label{Eq_iterate}
\end{equation}

The convergence of $\delta^n r_1$ series can be proved under two conditions:\\
(i) All quantities vary slowly along the space, namely $|P^\prime/P^2|\ll 1$ is satisfied for any quantity $P(x)$.\\
(ii) $\delta^n r_1$ is much smaller compared to the other roots of the polynomial. 

If we denote the other roots as $\delta^n r_j$ and $j\neq 1$, the polynomial coefficients under the second condition can be rewritten as $\delta ^n H_0=-\delta^n r_1 \prod_{j\neq 1}(-\delta^n r_j)$ and $\delta^n H_1\simeq \prod_{j\neq 1}(-\delta^n r_j)$. Thus, we obtain
\begin{equation}
    \delta^n r_1=-\delta^n H_0/\delta^n H_1.
\end{equation}
Expanding the coefficients on the right-hand side using Eq. (\ref{Eq_iterate}) and use the first condition and Eq. (\ref{Eq_IntPart}) to simplify, we find an iterative relation for $\delta^n r_1$ as 
\begin{equation}
    \delta^{n}r_1\simeq -\dfrac{\delta^{n-1}H_{2}}{\delta^{n-1}H_{1}}\delta^{n-1}r_1^{\prime}.
    \label{Eq_iterater1}
\end{equation}
Applying the slow-varying condition, we have $|\delta^n r_1|\ll \left|\frac{\delta^{n-1}H_2}{\delta^{n-1}H_1}\delta^{n-1}r_1^2\right|$. The term $\left|\frac{\delta^{n-1}H_2}{\delta^{n-1} H_1}\right|$ can be expressed as $\left|\sum_{l\neq 1} \frac{1}{\delta^{n-1}r_j}\right|$, which is much smaller than $\left|\frac{1}{\delta^{n-1}r_1}\right|$ because of the second condition.
Therefore, we have $|\delta^n r_1|\ll |\delta^{n-1} r_1|$, indicating that the $\delta^n r_1$ series ultimately converges to zero as the iteration order $n$ increases.

\subsection{Relation to the parametric dispersion relation }


To understand the underlying physics described by our amplitude equation and its solutions, we need to connect coefficients $H_p$ to the physical quantities we are familiar with, such as the growth rate $\gamma$, the linear dispersion relation $\varepsilon_u$ and the coupling coefficients $\alpha$. Results in terms of these quantities also serves as an analytical verification of our approach in comparison to the previous studies.

Revisiting the Eqs. (\ref{Eq_CoupWave1}, \ref{Eq_CoupWave2}), if we consider small neighborhoods around each point and assume that the medium is approximately homogeneous within these regions, the coupling wave equations can be linearized to the familiar forms $\varepsilon_u \Phi_u=\alpha \Phi_1$ and $\varepsilon_1 \Phi_1=\alpha^*  \Phi_u$, which directly yield the parametric dispersion relation $\varepsilon_u\varepsilon_1-\alpha\alpha^*=0$. The correspondence between coefficients $b_n, c_n, a_m, a_m^*$ in the coupling wave equations and the terms in the parametric dispersion relation is
\begin{equation}
\begin{split}
    \varepsilon_u(k;x)=&\sum_{n=0}^N b_n(x) (ik)^n,\\
    \varepsilon_1(k;x)=&\sum_{n=0}^{N_1}c_n(x)(ik)^n,\\
    \alpha(k;x)=&\sum_{m=0}^M a_m(x)(ik)^m,\\
    \alpha^*(k;x)=&\sum_{m=0}^M a_m^*(x)(ik)^m.
\end{split}
\label{Eq_KtoEps}
\end{equation}

On the other hand, for weak inhomogeneity, we can neglect all the parameter gradients (e.g., $a_m^\prime$, $b_n^\prime$, etc.) as shown in Eqs. (\ref{Eq_Mbj}, \ref{Eq_V1ajp}). Each component in Eq. (\ref{Eq_HpDet}) can then be greatly simplified, such as $\boldsymbol{M}_b\simeq \boldsymbol{M}_{b0}$, $\boldsymbol{M}_{a^*}\simeq \boldsymbol{M}_{a^*0}$, and $V_{1aj}\simeq V_{1am}=\sum_{n=0}^M a_n\left(\phi_1^{(n)}e^{i\Psi_0}\right)^{(m)}$, $V_{1cj}\simeq V_{1cm}=\sum_{n=0}^{N_1} c_n\left(\phi_1^{(n)}e^{i\Psi_0}\right)^{(m)}.$ Leveraging the triangular-like structure of matrix $\boldsymbol{M}_{b0}$ and $\boldsymbol{M}_{a^*0}$, we have 
\begin{equation}
    H_p=(-1)^N\mathrm{det}|\bar{\boldsymbol{M}}_{ab}|\left[\sum_{m=0}^{N}b_{m}V_{1cmp}-\sum_{m=0}^{M}a_{m}^{*}V_{1amp}\right].
    \label{Eq_HpWInhom}
\end{equation}

The common factor $(-1)^N\mathrm{det}|\boldsymbol{\bar{M}}_{ab}|$ in Eq. (\ref{Eq_HpWInhom}) of $H_p$ can be omitted, where $\boldsymbol{\bar{M}}_{ab}$ is a $(\text{M}+\text{N})*(\text{M}+\text{N})$ coefficient matrix given by 
\begin{equation}
    \boldsymbol{\bar{M}}_{ab}=\left(\begin{array}{cccccc}
b_{0} & \cdots & \cdots & b_{N}\\
 & \ddots & \ddots & \ddots & \ddots\\
 &  & b_{0} & \cdots & \cdots & b_{N}\\
a_{0}^{*} & \cdots & a_{M}^{*}\\
 & \ddots & \ddots & \ddots\\
 &  & \ddots & \ddots & \ddots\\
 &  &  & a_{0}^{*} & \cdots & a_{M}^{*}
\end{array}\right).
\end{equation}

Ignoring the parameter gradients also helps in simplifying $V_{1amp}$ and $V_{1cmp}$ in Eq. (\ref{Eq_HpWInhom}). Using  $\left[\exp(i\Psi_1)\right]^{(m)}\simeq (ik_1)^me^{i\Psi_1}$, we can approximate Eq. (\ref{Eq_V1ajp}) as


\begin{equation}
    V_{1amp}\simeq e^{i\Psi_1+i\Psi_0}\sum_{n=0}^{N_1} a_n\sum_{m_{a}=0}^{m}C^m_{m_a}C^{n+m_a}_p(ik_{1})^{n-p+m_{a}}(ik_{0})^{m-m_{a}}.
\end{equation}

Next, we handle the sum over $m_a$. After expanding the binomial coefficient $C^{n+m_a}_p$ to $\sum_{l=0}^p C^n_{p-l}C^{m_a}_l$ using the Vandermonde's  identity and subsequently transform $C^m_{m_a}C^{m_a}_l$ into $C_{m-l}^{m_a-l}C_l^m$, the sum over $m_a$ turns out to be a binary expansion of $(ik_0+ik_1)^{m-l}$. Denote $k_{s}\equiv k_0+k_1$, we obtain

\begin{equation}
V_{1amp}\simeq e^{i\Psi_1+i\Psi_0}\sum_{n=0}^{N_1} a_n\sum_{l=0}^p
C^n_{p-l}(ik_{1})^{n-(p-l)}C^m_l(ik_{s})^{m-l}.
\end{equation}
Substituting simplified $V_{1amp}$ and $V_{1cmp}$ into Eq. (\ref{Eq_HpWInhom}), and using Eqs. (\ref{Eq_KtoEps}) to replace the parameters like $a_n$ with the terms of the parametric dispersion relation like $\alpha$, the coefficient $H_p$ is finally
\begin{equation}
    H_p=(-i)^p\sum_{l=0}^p\dfrac{\dfrac{\partial^{p-l}}{\partial k_1^{p-l}}\dfrac{\partial^l}{\partial k_s^l}}{l!(p-l)!}\left[\varepsilon_1(k_1)\varepsilon_u(k_s)-\alpha(k_1)\alpha^*(k_s)\right],
    \label{Eq_HpFinCW}
\end{equation}
or equivalently
\begin{equation}
    H_p=\dfrac{(-i)^p}{p!}\dfrac{\partial^p}{\partial k_1^p}\left[\varepsilon_1(k_1)\varepsilon_u(k_s)-\alpha(k_1)\alpha^*(k_s)\right]
\end{equation}
using the relation $\frac{\partial}{\partial (k_0+k_1)}=\frac{\partial}{\partial k_1}$.

This result resembles a Taylor expansion on the local parametric dispersion relation with respect to the intrinsic wave number $k_1$. This similarity arises because there is a method derive the amplitude equation merely by Taylor expanding the parametric dispersion relation, albeit not rigorously: Assume the parametric dispersion relation universally satisfies $\varepsilon_{NL}=0$ , and the resonant wave '1' oscillates with an exponential rate as $ik_1+r_1$, where the intrinsic wave number $k_1$ is the primary component and  $r_1$ is the minor modification induced by inhomogeneity. Using $|r_1|\ll |k_1|$, we can expand $0=\varepsilon_{NL}(ik_1+r_1)=\sum_{p=0}^\infty \frac{(-i)^p}{p!}\frac{\partial^p \varepsilon_{NL}}{\partial k_1^p} r_1^p$, which corresponds to the characteristic polynomial $\sum_p H_pr_1^p=0$. 


\section{Effect of wave number mismatch on parametric instability\label{sec_3}}

Previous theories have left the wave number of the quasi-mode in non-resonant decay unresolved, a key focus here. In Sec. \ref{sec_2}, our approach avoids directly assuming the quasi-mode wave vector by eliminating $\phi_u$ in Eq. (\ref{Eq_det}), bypassing related controversies. Here, we deduce the appropriate quasi-mode wave vector by comprehensively calculating the wave vector mismatch and amplification factor for various PI types, especially the non-resonant type.

To exclude the effect of parameter gradients, this section focuses on very weak inhomogeneity. In this condition, performing the first iteration suffices to solve the amplitude equation, yielding $R_1\simeq r_1$. If $r_1$ is small, it can be analytically expressed with the roots of the characteristic polynomial truncated at the first and second orders as 
\begin{equation}
    r_1^{\text{od1}}=-\dfrac{H_0}{H_1},
    \label{eq_r1od1}
\end{equation}
\begin{equation}
    r_1^{\text{od2}}=-\dfrac{H_1}{2H_2}\pm\sqrt{\dfrac{H_1^2}{4H_2^2}-\dfrac{H_0}{H_2}},
    \label{eq_r1od2}
\end{equation}
respectively.

\subsection{Resonant type}

For resonant type PI, mode 'u' is a resonant wave with one or several intrinsic wave numbers in the medium, denoted as $k_{rj}$, which is the real roots of the linear dispersion relation $\varepsilon_u(k_{rj}(x);x)=0$. For each $k_{rj}$, the wave number mismatch is $\kappa_j =k_{rj}-k_0-k_1$, and the matching point $x_{jm}$ satisfies $\kappa_j(x_{jm})=0$.

Let us begin with the coefficients $H_p$ of the amplitude equation. Near the $j$th matching point $x_{jm}$ where $\kappa_j(x_{jm})=0$, the wave number mismatch $\kappa_j$ is small. Thus, we approximate $\varepsilon_u(k_s)\simeq -\frac{\partial \varepsilon_u}{\partial k_{rj}}\kappa_j$ and $\frac{\partial \varepsilon_u}{\partial k_s}\simeq\frac{\partial \varepsilon_u}{\partial k_{rj}}-\frac{\partial^2 \varepsilon_u}{\partial k_{rj}^2}\kappa_j$. The coefficients simplify to

\begin{equation}
\begin{split}
    H_0=&-\alpha\alpha^*,\\
    H_1\simeq &i\kappa_j\frac{\partial \varepsilon_u}{\partial k_{rj}}\dfrac{\partial \varepsilon_1}{\partial k_1},\\
    H_2\simeq &-\dfrac{\partial \varepsilon_u}{\partial k_{rj}}\dfrac{\partial \varepsilon_1}{\partial k_1},\\
    H_3\simeq &\dfrac{1}{2}\dfrac{\partial^2 \varepsilon_u}{\partial k_{rj}^2}\dfrac{\partial \varepsilon_1}{\partial k_{1}}+\dfrac{1}{2}\dfrac{\partial \varepsilon_u}{\partial k_{rj}}\dfrac{\partial^2 \varepsilon_1}{\partial k_{1}^2}.
\end{split}
\label{Eq_H012Re}
\end{equation}
Since $H_1$ is small, $r_1^{\text{od}1}$ leads to large value, making it unsuitable for describing the growth rate. Instead, $H_2\propto v_{gu}v_{g1}$ is not small, so $r_1^{\text{od}2}$ is generally valid unless one the child waves do not propagate. 

Substituting these coefficients into $r_1^{\text{od2}}$, we obtain the spatial growth rate $\mathrm{Re}(r_1^{\text{od}2})$ 
\begin{equation}
    \mathrm{Re}(r_1^{\text{od}2})=\sqrt{-\Lambda_j-\dfrac{1}{4}\kappa_j^2},
    \label{Eq_r1Re}
\end{equation}
where $\Lambda_j\equiv \alpha\alpha^*\left(\frac{\partial \varepsilon_1}{\partial k_1}\frac{\partial \varepsilon_u}{\partial k_{uj}}\right)^{-1}.$

The growth rate is consistent with Ref. \cite{rosenbluthPhys.Rev.Lett.1972}, where $-\Lambda_j$ takes the form of $\frac{\gamma_0^2} {v_{guj}v_{g1}}$. When the wave numbers match, i.e., $\kappa_j=0$, the spatial growth rate described by Eq. (\ref{Eq_r1Re}) can reduce to the growth rate reported in Eq. (\ref{Eq_GamRe}). 

Eq. (\ref{Eq_r1Re}) clearly illustrates the mechanism by which wave number mismatch inhibits the resonant type PI: The PI growth rate is maximal at the matching point. As child waves propagate away from this point, the growth rate diminishes due to the increasing wave number mismatch and ultimately vanishes when the wave reaches the turning point $ x_t=\pm 2\sqrt{-\Lambda_j}/|\kappa_j^\prime|$.

The derivation of amplification factor $A_{\text{inhom}}\equiv\int_{-\infty}^\infty \mathrm{Re}(r_1^{\text{od}2})dx$ for the resonant type also generally follows the result of Ref.\cite{rosenbluthPhys.Rev.Lett.1972}. Our analysis differs by considering multiple eigenmodes of $k_{rj}$, leading to more than one matching point. The total amplification factor is the sum of contributions from each matching point

\begin{equation}
    A_{\mathrm{inhom}}=-\sum_j\dfrac{\pi\Lambda_j}{|\kappa_j^\prime|}.
    \label{Eq_AinhomReNew}
\end{equation}

For the child wave '1' propagating in the positive $x$ direction ($v_{g1}>0$, default in this paper), ensuring wave '1' amplification requires $\Lambda_j<0$ in Eq.(\ref{Eq_AinhomReNew}). Given that $-\Lambda_j=\frac{\gamma_0^2}{v_{g1}v_{guj}}$, only forward-scattered child waves achieve amplification for resonant type PI, while back-scattered child waves ($\Lambda_j>0$) do not. 


\subsection{Non-resonant type}

The mode 'u' here is off resonance, meaning its linear dispersion relation $\varepsilon_u$ is far from zero. The coefficients $H_p$s for non-resonant type PI are
\begin{equation}
\begin{split}
    H_0=&-\alpha\alpha^*,\\
    H_1\simeq&-i\varepsilon_u\dfrac{\partial\varepsilon_1}{\partial k_1},\\
    H_2\simeq&-\dfrac{1}{2}\varepsilon_u\dfrac{\partial^2 \varepsilon_1}{\partial k_1^2}-\dfrac{\partial \varepsilon_u}{\partial k_s}\dfrac{\partial \varepsilon_1}{\partial k_1}.
\end{split}
\end{equation}
Unlike in the resonant type PI, $H_1$ is no longer a small quantity, thus the characteristic root $r_1^{\text{od}1}$ is generally valid unless very strong PI coupling occurs ($|H_0|\gg |H_1|$). The spatial growth rate of non-resonant type PI is
\begin{equation}
    \mathrm{Re}(r_1^{\text{od}1})\simeq \dfrac{\mathrm{Im}\left[-\alpha\alpha^*/\varepsilon_u\right]}{\partial \varepsilon_1/\partial k_1}.
    \label{Eq_r1Nr}
\end{equation}
This meets the spatial growth rate $\gamma/v_{g1}$ derived for the non-resonant type PI in previous studies\cite{liuPhys.Rep.-Rev.Sect.Phys.Lett.1985}, where $\gamma$ has been shown by Eq. (\ref{Eq_GamNr}). 

Just as the wave number mismatch $\kappa_j$ terms are obtained by writing $\varepsilon_u(k_s)\simeq -\frac{\partial \varepsilon_u}{\partial k_s}\kappa_j$ in $\mathrm{Re}(r_1^{\text{od}2})$ for the resonant type, we may also find these key terms here from $\varepsilon_u(k_s)$ in $\mathrm{Re}(r_1^{\text{od}1})$. Noting that the function $\varepsilon_u(k)$ is a Nth-degree polynomial as shown by Eq. (\ref{Eq_KtoEps}), we can rewrite $\varepsilon_u(k_s)$ as  %
\begin{equation}
    \varepsilon_u(k_s)=b_N(-i)^N \prod_{j=1}^N \left(k_{cj}-k_0-k_1\right)
    \label{eq_epstoprod}
\end{equation}
where $k_{cj}$ is the $j$th root of polynomial $\varepsilon_u(k)=0$. All the $k_{cj}$ should be complex because any real root would imply that the 'u' mode is a resonant wave. 

Recalling that the core characteristic of wave number mismatch $\kappa_j$ is its increase leading to a decrease in spatial growth rate, we difine the wave number mismatch $\kappa_j\equiv k_{cj}-k_0-k_1$ since it meets this characteristic. This definition also shares a common point with the definition $\kappa_j\equiv k_{rj}-k_0-k_1$ of the resonant type, as both $k_{rj}$ and $k_{cj}$ are the roots of wave number of mode 'u'. 

For the amplification factor $A_{\text{inhom}}\equiv\int_{-\infty}^\infty \mathrm{Re}(r_1^{\text{od}1})dx$ for the non-resonant type PI, we prefer to write the integral as
\begin{equation}
    A_{\text{inhom}}=\dfrac{1}{2}\int_{-\infty}^{+\infty}r_1^{\text{od1}}dx+\dfrac{1}{2}\int_{-\infty}^{+\infty}r_1^{\text{od1}*}dx,
\end{equation}
where $r_1^{\text{od1}*}$ is the conjugate of $r_1^{\text{od1}}$. 

Let's begin by handling the first integral. In contrast to the resonant type, the integration interval extends to $\left(-\infty,+\infty\right)$ and is no longer limited by the turning points. By extending the integral domain to the complex plane, we can write it as
\begin{equation}
    \int_{-\infty}^{+\infty}r_{1}dx=\dfrac{i^{N+1}\alpha\alpha^{*}}{\frac{\partial\varepsilon_{1}}{\partial k_{1}}b_{N}}\left(\oint_{C}\dfrac{dz}{\prod_{j}\kappa_{j}}-\lim_{R\to\infty}\int_{0}^{\pi}\dfrac{Rd\theta}{\prod_{j}|\kappa_{j}|e^{i\theta_{j}}}\right),
    \label{eq_Intr1}
\end{equation}
where the closed integral path $C$ encloses the upper half of the complex plane. Terms other than $\kappa_j$ are assumed to be approximately independent of $x$ since we focus primarily on effect of the wave number mismatch. To ensure the convergence at infinity, $\kappa_j(z)$ is defined as a monotonically continuous function that grows at least at a linear rate or faster. Therefore, we can deal the first term in Eq. (\ref{eq_Intr1}) with the Residue Theorem, and the second integral, where $\kappa_j$ is expressed in polar coordinates as $|\kappa_j|e^{i\theta_j}$, is approximately neglected.

Denote the singular points as $z_{j\pm}$, which also serves as the matching points as they satisfy $\kappa_j(z_{j\pm})=0$. The sign in subscript of $z_{j\pm}$ indicates whether the point is in the upper or lower part of the complex plane. It results in

\begin{equation}
    \int_{-\infty}^{+\infty} r_1 dx=\sum_{j+} \dfrac{-2\pi\alpha\alpha^*}{\dfrac{\partial \varepsilon_1}{\partial k_1}\dfrac{\partial \varepsilon_u}{\partial k_{cj+}}\kappa_{j+}^\prime},
\end{equation}

where $\frac{\partial \varepsilon_u}{\partial k_{cj+}}=\left.\frac{\partial \varepsilon_u(k)}{\partial k}\right|_{k=k_{cj+}}=-b_N(-i)^N\prod_{j\neq j_+}\kappa_j(z_{j+})$ using Eq. (\ref{eq_epstoprod}). Performing a similar derivation on the conjugate integral and sum over the contributions of all matching points, the total amplification factor becomes
\begin{equation}
    A_{\mathrm{inhom}}=\sum_{j+}\dfrac{\pi\Lambda_{j+}}{\kappa_{j+}^\prime}-\sum_{j-}\dfrac{\pi\Lambda^*_{j-}}{\kappa_{j-}^{*\prime}}.
    \label{Eq_AinhomNrNew1}
\end{equation}
The quantity $\Lambda_{j\pm}$ follows its definition in the resonant type PI. 

Each $j$ term in Eq. (\ref{Eq_AinhomNrNew1}) resembles the form of the resonant type PI in Eq. (\ref{Eq_AinhomReNew}). One fundamental difference is that $k_{rj}$ for resonant type is real while $k_{cj}$ for non-resonant type is complex. This distinction can be unified with the framework that both $k_{rj}$ and $k_{cj}$ are roots of the linear dispersion equation $\varepsilon_u(k)=0$. Roots on the real axis ($\mathrm{Im}(k_{j})=0$) correspond to resonant type PI, whereas roots off the real axis ($\mathrm{Im}(k_{j})\neq 0$) correspond to non-resonant type PI.

Another difference is that the non-resonant amplitude factor may include contributions from both forward-scattered and backward-scattered matching points. To illustrate this, consider a forward-scattered scenario ($\Lambda_j<0$, thus $v_{guj}>0$ and $\text{Im}(k_{cj})>0$). Assuming a linear-varying $\kappa_j$, the matching point is then at $z_j=-\kappa_j(0)/\kappa_j^\prime$. Because $\text{Im}(\kappa_j(0))=\text{Im}(k_{cj})>0$, whether the matching point is in upper or lower half of the complex plane depends solely on the sign of $\kappa_j^\prime$. For $\kappa_j^\prime >0$, this matching point's contribution is $-\pi\Lambda_j^*/\kappa_j^{*\prime}$, while for $\kappa_j^\prime <0$ the contribution is $\pi\Lambda_j/\kappa_j^{\prime}$.
Conversely, it is easy to find that a backward-scattered matching point ($\Lambda_j>0$) also contributes to the amplification as $\pi\Lambda_j/\kappa_j^{\prime}$ for $\kappa_j^\prime>0$ and $-\pi\Lambda_j^*/\kappa_j^{*\prime}$ for $\kappa_j<0$. Based on the above analysis, we can rewrite Eqs. (\ref{Eq_AinhomNrNew1}) to distinguish forward and backward-scattered types as
\begin{equation}
     A_{\text{inhom}}=-\sum_{j\in \text{Forward}}\text{sign}(\kappa_{rj}^\prime)\text{Re}\left(\dfrac{\pi\Lambda_{j}}{\kappa_{j}^\prime}\right)+\sum_{j\in \text{Backward}}\text{sign}(\kappa_{rj}^\prime)\text{Re}\left(\dfrac{\pi\Lambda_{j}}{\kappa_{j}^\prime}\right).
     \label{Eq_AinhomNrNew2}
\end{equation}

From Eqs. (\ref{Eq_AinhomReNew}) and (\ref{Eq_AinhomNrNew2}), we conclude that the quasi-mode's intrinsic is a complex root of the linear dispersion equation, not arbitrarily chosen. Quasi-modes, like the resonant wave, inherently exist within the mode structure of media, though they are harder to excite due to the strong damping. 
For non-resonant type PI, the wave number mismatch is $\kappa_j=k_{cj}-k_0-k_1$. Its matching points are off the real axis and practically unreachable. Instead, we locate a point on the real axis with maximum spatial growth rate where $|\kappa_j|$ is minimized. This quasi-matching point, $x_{j}$, typically satisfies $\text{Re}[\kappa_j(x_{j})]=0$. The closer $x_{j}$ is to the matching point $z_j$, the greater its contribution to the amplification factor.


\subsection{Intermediate type}

The physical images of resonant and non-resonant types seem to merge based on the above analysis, but a significant gap remains for the backward-scattered PI. It is not supported by the resonant PI but can occur in the non-resonant type. To bridge this gap, let's examine an intermediate type between resonant and non-resonant types.

Since $\mathrm{Im}(k_{rj})=0$ represents a resonant type, and $|\mathrm{Im}(k_{cj})|\gg 1$ represents a quasi-mode and non-resonant type, an intermediate type refers to a nearly resonant mode 'u' with a small damping. By expanding the wave number into real and imaginary parts $k_j=k_{rj}+ik_{ij}$, where $k_{ij}$ is small, we can approximately write $r_1^{\text{od}2}$ as 

\begin{equation}
    r_1^{\text{od}2}\simeq\dfrac{i\kappa_j-k_{ij}}{2}\pm\sqrt{-\Lambda_{rj}-\dfrac{1}{4}\kappa_j^2-\dfrac{1}{4}k_{ij}^2-\dfrac{i}{2}\kappa_jk_{ij}},
    \label{eq_r1Intm}
\end{equation}
where $\Lambda_{rj}\equiv \alpha\alpha^*\left(\frac{\partial \varepsilon_1}{\partial k_1}\frac{\partial \varepsilon_u}{\partial k_{rj}}\right)^{-1}$. 

\subsubsection{Forward scattering}
The forward-scattered PI implies $\Lambda_{rj}<0$ and $k_{ij}>0$, hence the primary real part of Eq. (\ref{eq_r1Intm}) is $\sqrt{-\Lambda_{rj}}$ at the matching point and 0 outside the turning points. The introduction of small quantity $k_{ij}$ slightly modifies this. Using approximations $k_{ij}^2\ll -\Lambda_{rj}$ near the matching points and $|\kappa_j|\gg -\Lambda_{rj}$ outside the turning point, the PI spatial growth rate is
\begin{equation}
    \mathrm{Re}(r_{1}^{\mathrm{od}2})\simeq\begin{cases}
-\dfrac{1}{2}k_{ij}+\sqrt{-\Lambda_{rj}-\dfrac{1}{4}\kappa_{j}^{2}} & \text{near the matching point}\\
-\dfrac{1}{2}\dfrac{k_{ij}\Lambda_{rj}}{\kappa_{j}^{2}} & \text{outside the turning point}
\end{cases}
\end{equation}
The corresponding amplification factor can be obtained by piecewise integrating the upper expression within the turning points $[-x_t,x_t]$, and integrating the lower expression within intervals $(-\infty,-x_t]$ and $[x_t,+\infty)$. It can be found that the inhibiting effect of $k_{ij}$ near the matching point will cancel with its enhancing effect outside the turning point. As a result, the contribution of $j$th matching point to the amplification factor is
\begin{equation}
    A_{\text{inhom}}|_j=\frac{-\pi\Lambda_j}{|\kappa_j^\prime|},
    \label{Eq_AinhomItFw}
\end{equation}
in the same form with Eqs. (\ref{Eq_AinhomReNew}) and (\ref{Eq_AinhomNrNew1}).

\subsubsection{Backward scattering}
The backward-scattered PI implies $\Lambda_j>0$ and $k_{ij}<0$. Since its spatial growth rate for resonant type is completely zero, introducing a small $k_{ij}$ cannot trigger a large growth rate. Using the approximation $k_{ij}^2\ll \Lambda_j$ near the matching point, we can write Eq. (\ref{eq_r1Intm}) as
\begin{equation}
    \mathrm{Re}(r_1^{\text{od}2})\simeq -\dfrac{1}{2}k_{ij}+\dfrac{|\kappa_j|k_{ij}}{2\sqrt{\kappa_j^2+4\Lambda_j}}.
    \label{Eq_r1ItBW}
\end{equation}
The growth rate here exhibits a linearly increasing dependence on the damping rate $k_{ij}$ of mode 'u', implying that a backward child wave with stronger damping might abnormally trigger PI amplification. Nevertheless, the growth rate still follows the common pattern that PI is inhibited by the wave number mismatch. As $\kappa_j$ increases from zero to infinity, the growth rate will decrease monotonically from $-\frac{1}{2}k_{ij}$ to zero.

The corresponding amplification factor is calculated by piecewise integrating Eq. (\ref{Eq_r1ItBW}) over intervals $(-\infty,x_{mj}]$ and $[x_{mj},+\infty)$, resulting in
\begin{equation}
    A_{\text{inhom}}=\sum_j\dfrac{-2k_{ij}\sqrt{\Lambda_j}}{|\kappa_j^\prime|}.
    \label{Eq_AinhomItBw}
\end{equation}
This expression bridges the gap between the no amplification result of the resonant type and a finite amplification described by Eq. (\ref{Eq_AinhomNrNew2}). It predicts that when $-k_{ij}$ is sufficiently large, such as $-k_{ij}\ge \frac{\pi}{2}\sqrt{\Lambda_j}$, this intermediate type PI will completely transition to a non-resonant type.

Concluding from Eqs. (\ref{Eq_AinhomReNew}),  (\ref{Eq_AinhomNrNew2}),  (\ref{Eq_AinhomItFw}) and (\ref{Eq_AinhomItFw}), we observe that $\pm \pi \Lambda_j/|\kappa_j^\prime|$ serves a unified formula for amplification factors across various PI types. However, this unification does not apply to backward-scattered PI with resonant or nearly resonant child wave 'u'. 

\subsection{Numerical Examples}

We perform subsequent numerical studies for two aims. First is to intuitively show the difference of mismatching mechanisms between the resonant type and non-resonant type PI, as well as that between forward scattering and backward scattering. Second is to use these specific results to verify the physics based on our analytical analysis. 

Consider a simple set of coupling wave equations where the highest order is 2: $\text{N}=\text{N}_1=2$ and $\text{M}=0$. The coefficients $b_2, c_2=1$; $b_1, c_1=0$ ; $b_0=K_u^2(x)$ and $c_0=K_1^2(x)$, resulting in the intrinsic wave numbers are $k_u=\pm K_u$ and $k_1=\pm K_1$, respectively. The pump wave oscillates with only one branch of intrinsic wave number $K_0(x)$, and we denote $K_s\equiv K_0+K_1$.  Therefore, the coupling wave equations in the form of Eqs. (\ref{Eq_CoupWave1}, \ref{Eq_CoupWave2}) are 
\begin{equation}
    \phi_u^{\prime\prime}+K_u^2(x)\phi_u=a_0 e^{i\Psi_0}\phi_1,
\end{equation}
\begin{equation}
    \phi_1^{\prime\prime}+K_1^2(x)\phi_1=a_0^*e^{-i\Psi_0} \phi_u.
\end{equation}

The specific value of the parameters are as follows: The intrinsic wave numbers are $K_0=9.0-0.01x$, $K_1=-9.0$, and $\mathrm{Re}(K_u)=4.0$.  The slow change of $K_0$ with respect to $x$ introduces a weak inhomogeneous scenario, where a forward-scattered matching point is found at $x=400$ satisfying $K_u-K_0-K_1=0$, and a backward-scattered matching point is found at $x=-400$ satisfying $-K_u-K_0-K_1=0$. Numerical calculations will be performed for different $\text{Im}(K_1)$ to simulate resonant, non-resonant and intermediate types of PI. The coupling coefficients are $a_0=a_0^*=2.0$. 

We will substitute these parameters into Eq. (\ref{eq_r1od1}) and Eq. (\ref{eq_r1od2}) to calculate growth rates for each point of the domain, and then integrate them point by point to obtain the amplification process of amplitude $\Phi_1$. These results serve as predictions of our analytical approach. In contrast to the theoretical predictions, numerical solutions are obtained using MATLAB's \textit{bvp4c} solver for differential equations. To ensure the triggered $\phi_1$ is mainly the $k_1=K_1$ branch of interest, the boundary conditions at $x_a$ and $x_b$ are set as
\begin{equation}
\begin{split}
    \phi_u(x_a)&=1,\\
    \phi_1(x_a)&=1,\\
    \phi_1^\prime(x_a)&=iK_1(x_a),\\
    \phi_1^\prime(x_b)&=iK_1(x_b)\phi_1(x_b).
\end{split}
\label{eq_boundary}
\end{equation}

\begin{figure}[htbp!]
    \centering
    \includegraphics[width=0.9\linewidth]{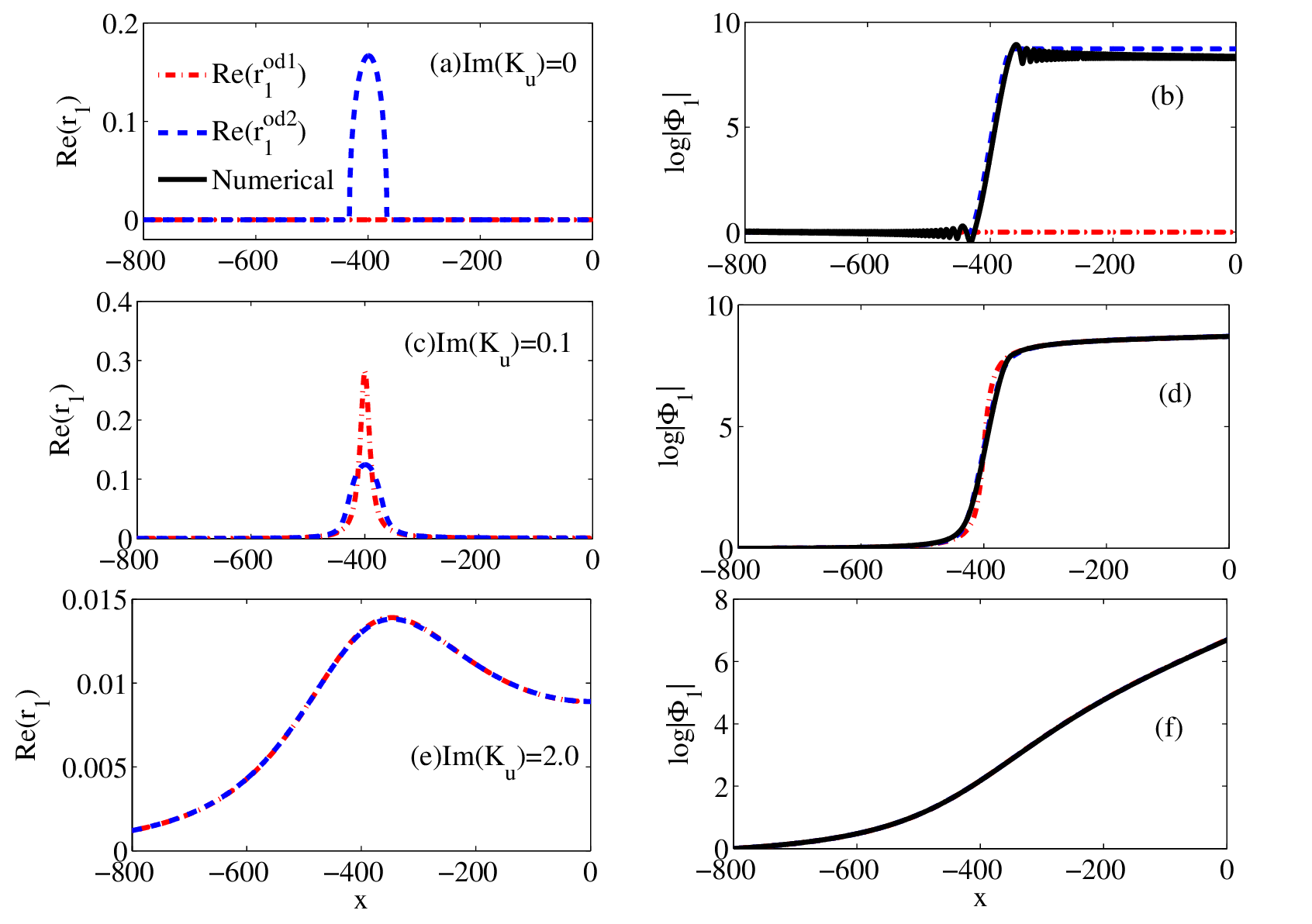}
    \caption{\label{fig_Fw}The spatial growth rate $\mathrm{Re(r_1)}$ (a, c, e) and the corresponding amplitude $\Phi_1$ of child wave '1' (b, d, f) as functions of $x$ in the vicinity of the forward-scattered matching point. Different rows of panels represent the resonant type (a, b), the intermediate type (c, d) and the non-resonant type (e, f) PIs, respectively. All the growth rates are solutions of the amplitude equation (\ref{Eq_AmpEq}). Red dot-dashed lines and blue dashed lines represent the results of Eq. (\ref{eq_r1od1}) and Eq. (\ref{eq_r1od2}), respectively. Black solid lines represent the direct numerical solutions of Eqs. (\ref{Eq_CoupWave1}) and (\ref{Eq_CoupWave2}).}
\end{figure}


Let's examine results near the forward-scattered matching point first. For the resonant type PI ($\text{Im}(K_u)=0$) shown by Fig. \ref{fig_Fw}(a), we observe a clear distinction between growth rates solved from amplitude equations truncated at first order as $\mathrm{Re}(r_1^{\text{od1}})$ and second order as $\text{Re}(r_1^{\text{od2}})$.  $\mathrm{Re}(r_1^{\text{od1}})$ (red dot-dashed line) indicates no PI growth, whereas $\text{Re}(r_1^{\text{od2}})$ is positive within the interval limited by the turning points. In Fig. \ref{fig_Fw}(b), the numerical result supports results of  $\text{Re}(r_1^{\text{od2}})$ rather than  $\text{Re}(r_1^{\text{od1}})$, following the previously mentioned idea that  $\text{Re}(r_1^{\text{od2}})$ is more suitable to predict the resonant type PI. 

Figs. \ref{fig_Fw}(c) and (d) show the result when $\mathrm{Im}(K_u)=0.1$, representing that the 'u' mode experiences slight damping and the intermediate type PI. In this scenario, $\mathrm{Re}(r_1^{\text{od1}})$  exhibits a sharp peak at the matching point because its denominator $H_1\propto \mathrm{Im}(K_u)$ is a small quantity. $\mathrm{Re}(r_1^{\text{od2}})$ is no longer strictly limited to the vicinity of the matching point, and has tails extending beyond the turning points. In Fig. \ref{fig_Fw}(d), the numerical PI growth again aligns with  $\mathrm{Re}(r_1^{\text{od2}})$. Nevertheless, the final amplification factor of $\mathrm{Re}(r_1^{\text{od1}})$, measured by the upper limit of the growth $\log [\Phi_1(x_b)/\Phi_1(x_a)]$, is nearly identical to the numerical result as Eqs. (\ref{Eq_AinhomNrNew1}) and (\ref{Eq_AinhomReNew}) predicted.

When $\mathrm{Im}(K_u)=2.0$, the 'u' child mode undergoes strong damping, indicating a scenario for non-resonant type PI. In this scenario, the growth rate $\text{Re}(r_1^{\text{od1}})$ now closely matches $\text{Re}(r_1^{\text{od1}})$ as shown by Fig. \ref{fig_Fw}(e). In Fig. \ref{fig_Fw}(f), the amplitude evolution predicted by both are supported by the numerical results, demonstrating the capability of $\text{Re}(r_1^{\text{od1}})$ and Eq. (\ref{eq_r1od1}) for the non-resonant type. Compared with the results of resonant and intermediate types, the maximum growth rates have significantly decreased, while the growth region has been much broadened as the turning points vanish.

\begin{figure}[htbp]
    \centering
    \includegraphics[width=0.9\linewidth]{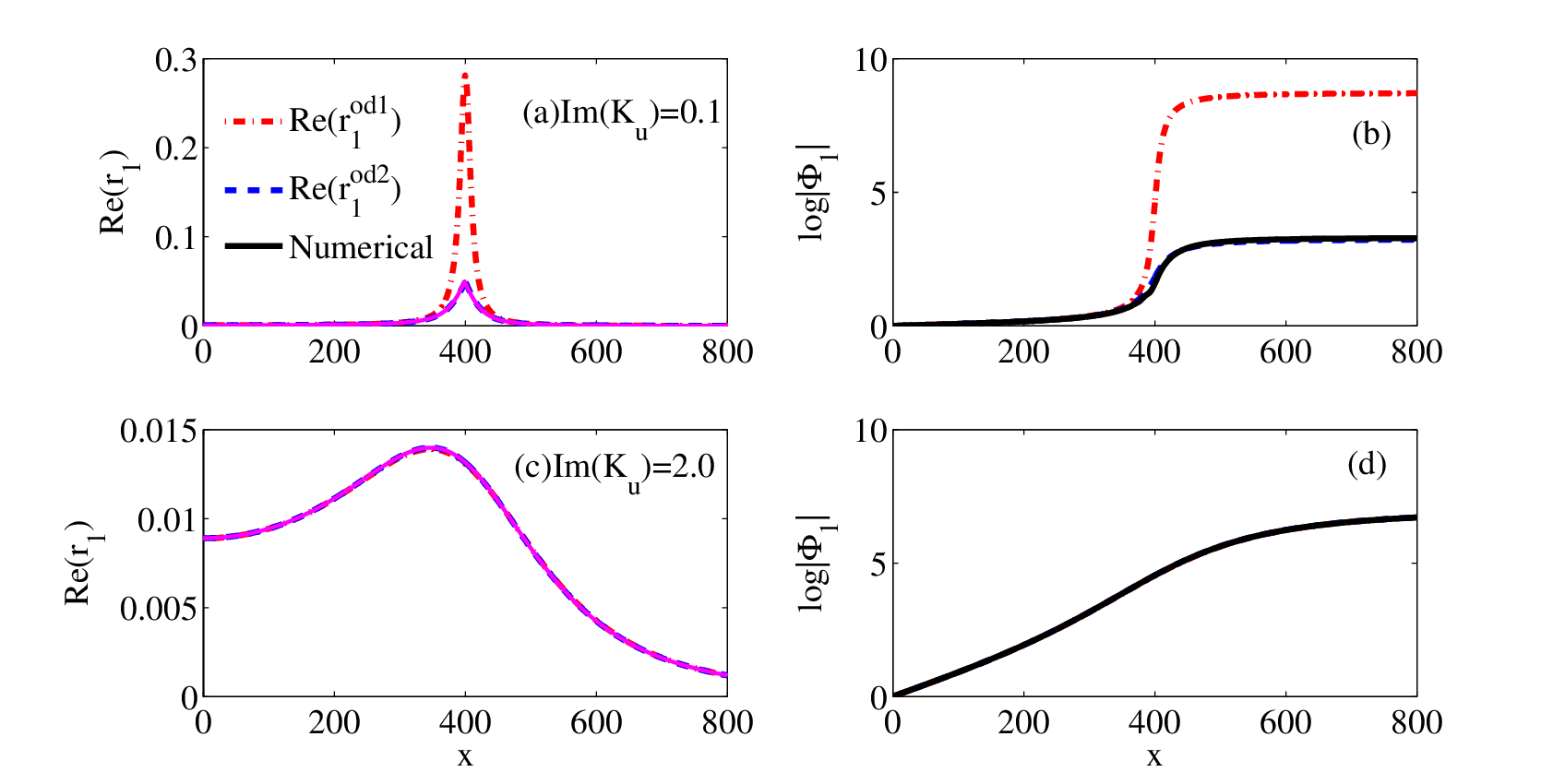}
    \caption{\label{fig_Bw}The spatial growth rate $\mathrm{Re(r_1)}$ (a, c) and the amplitude $\Phi_1$ of child wave '1' (b, d) as functions of $x$ in the vicinity of the backward-scattered matching point. Panels (a, b) illustrate the behavior of child wave '1' for intermediate type PI, while panels (c, d) depict that for non-resonant type PI. Other details of this figure follow those of Fig. \ref{fig_Fw}. }
\end{figure}

Fig. \ref{fig_Bw} presents the results in the vicinity of the backward-scattered matching point. For the resonant type PI, the numerical results, consistent with analytical predictions, exhibit no PI amplification across the domain and are thus omitted from the figure.  Figs. \ref{fig_Bw}(a) and (b) depict the intermediate type PI. A clear distinction between $\mathrm{Re}(r_1^{\text{od1}})$ and the more accurate $\mathrm{Re}(r_1^{\text{od2}})$ is observed, analogous to the forward scattering case. However, a crucial difference remains: the final amplification predicted by $\mathrm{Re}(r_1^{\text{od2}})$ for backward scattering is substantially lower and no longer consistent with the amplification predicted by $\mathrm{Re}(r_1^{\text{od1}})$. For the non-resonant type PI shown in Figs. \ref{fig_Bw}(c) and (d), $\mathrm{Re}(r_1^{\text{od1}})$ aligns well with the numerical results, reaffirming its predictive capability. Furthermore, the growth rate distribution in Fig. \ref{fig_Bw}(c) exhibits a mirror symmetry with Fig. \ref{fig_Fw}(e), suggesting that the growth rates of forward and backward scattering of non-resonant type PI are described by similar mathematical formulas. In summary, the backward scattering results corroborate the predictions of Eqs. (\ref{Eq_AinhomNrNew2}) and (\ref{Eq_AinhomItBw}), highlighting a pronounced distinction from forward scattering in the resonant and intermediate types, while exhibiting analogous behavior to forward scattering in the non-resonant type.
%


\begin{figure}[htbp]
    \centering
    \includegraphics[width=0.9\linewidth]{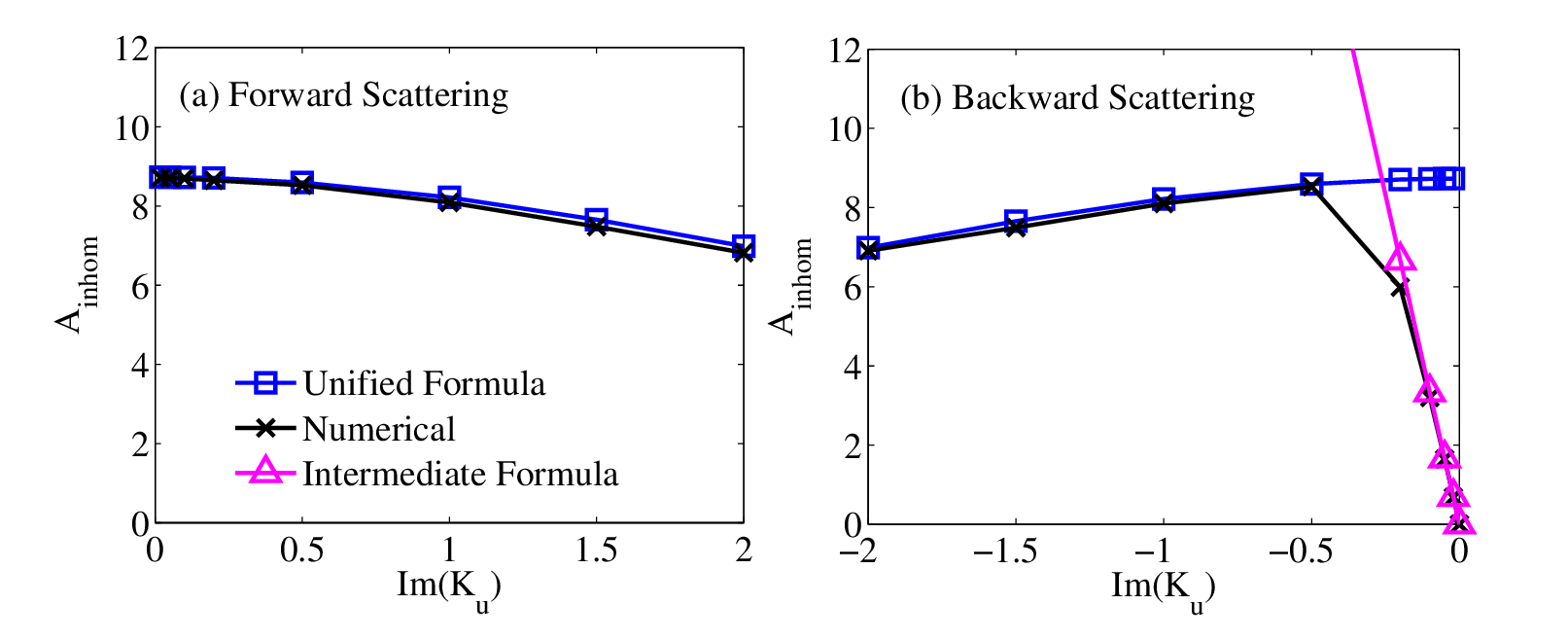}
    \caption{\label{fig_FwVSBw}Amplification factors contributed by (a) forward and (b) backward scattered matching points as functions of the linear spatial damping rate $\mathrm{Im}(K_u)$ of the 'u' mode. $\mathrm{Im}(K_u)$ equals to 0, closes to 0 and faraway from 0 represent the resonant, intermediate and non-resonant types of PI, respectively. Blue squares represent data results from the unified formula of amplification factor, given in Eqs. (\ref{Eq_AinhomReNew}), (\ref{Eq_AinhomNrNew2}) and (\ref{Eq_AinhomItFw}). Magenta triangles show data from Eq. (\ref{Eq_AinhomItBw}), illustrating the amplification factor of backward scattered intermediate type PI. Black crosses indicate the amplification factors derived from numerical solutions,  measured as $\log\frac{|\Phi_1(x_b)|}{|\Phi_1(x_a)|}$.}
\end{figure}

To validate the conclusion that $\pm \pi \Lambda_j/|\kappa_j^\prime|$ serves as an approximate unified formula for amplification factors and give an intuitive overview for these PI types, Fig. \ref{fig_FwVSBw} compares the amplification factors derived from the analytical formulas and numerical measurements as functions of $\text{Im}(K_u)$. For forward scattering, depicted by Fig. \ref{fig_FwVSBw}(a), we observe that the numerical results align with the analytical unified formula represented by the blue squares. For backward scattering, depicted by Fig. \ref{fig_FwVSBw}(b), the unified formula remains accurate in predicting the amplification factor for non-resonant types until $\text{Im}(K_u)$ exceeds approximately -0.5. As $\text{Im}(K_u)$ approaches zero, numerical results conform to Eq. (\ref{Eq_AinhomItBw}) and eventually decay to zero as the PI type transitions into a resonant one. 

\begin{figure}[H]
    \centering
    \includegraphics[width=0.9\linewidth]{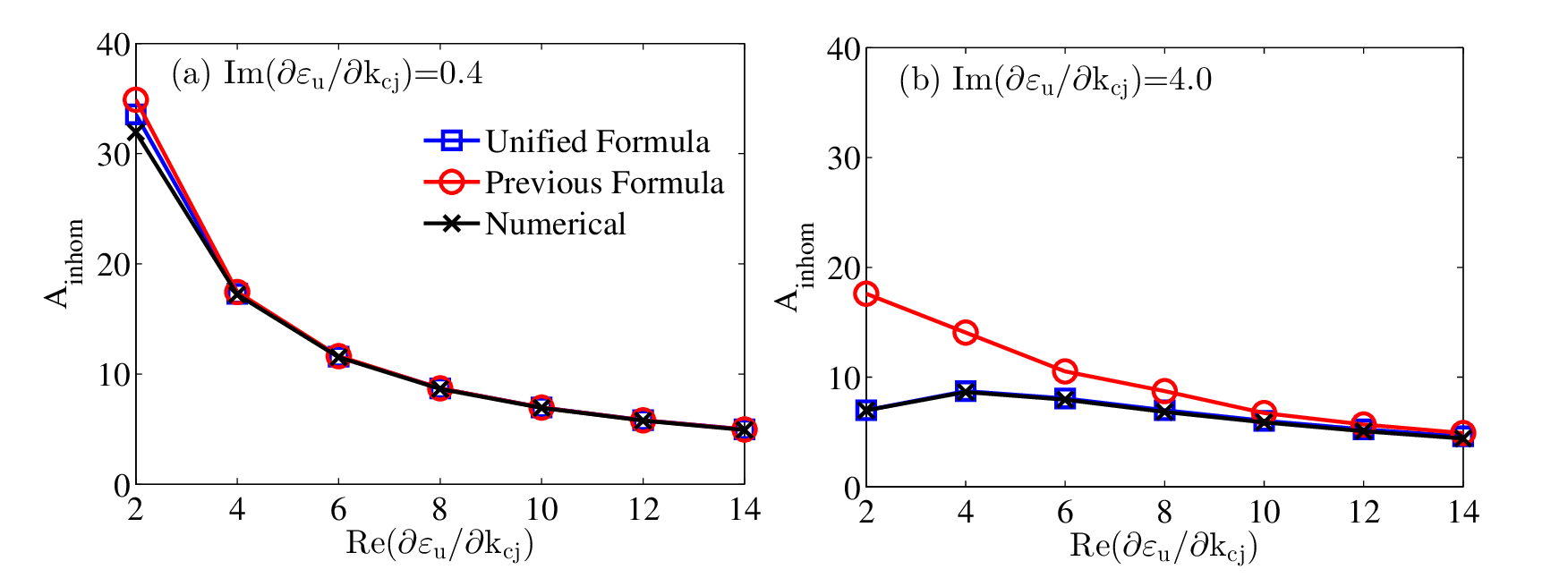}
    \caption{\label{fig_AinhomComp}Comparison between the amplification factors of non-resonant type PI described by unified formula in Eq. (\ref{Eq_AinhomNrNew2}) or Eq. (\ref{Eq_AinhomNr2}) derived by our approach (blue square) and the formula in Eq. (\ref{Eq_AinhomNr}) from previous studies (red circle). The comparison spans with respect to the real and imaginary part of the term $\frac{\partial \varepsilon_u}{\partial k_{cj}}$. Numerical results, depicted by black crosses, are included to validate these formulas.}
\end{figure}

Additionally, we compare our unified formula of amplification factor for non-resonant type PI with that from previous studies\cite{liuPhys.Rep.-Rev.Sect.Phys.Lett.1985,cesarioNucl.Fusion2006,zhaoPhys.Plasmas2013} derived using a presumed $k_u$. To facilitate a better comparison, we rewrite $-\pi\Lambda_j/|\kappa_j^\prime|$ in Eq. (\ref{Eq_AinhomNrNew2}) in a form similar to Eq. (\ref{Eq_AinhomNr}) as 
\begin{equation}
    A_{\mathrm{inhom}}|_j=\dfrac{\pi\gamma\mathrm{Im}\left(\varepsilon_u\right)}{v_{g1}}\mathrm{Re}\left[\left(\dfrac{\partial \varepsilon_u}{\partial k_{cj}}|\kappa_{j}^\prime|\right)^{-1}\right].
    \label{Eq_AinhomNr2}
\end{equation}
Except for the definition discrepancy that the preset $k_u$ is not exactly the complex intrinsic wave number $k_{cj}$ , the crucial difference between Eq. (\ref{Eq_AinhomNr2}) and Eq. (\ref{Eq_AinhomNr}) lies in the sequence of taking the real part and calculating the reciprocal of the term $\frac{\partial \varepsilon_u}{\partial k_{cj}}$. Assuming the preset $k_{u}$ is equivalent to $k_{cj}$, this difference becomes significant when the imaginary part of the $\frac{\partial \varepsilon_u}{\partial k_{cj}}$ dominates. In Fig. \ref{fig_AinhomComp},  we compare the two formulas of amplification factors with respect to the real and imaginary part of $\frac{\partial \varepsilon_u}{\partial k_{cj}}$. It is evident that the formula of previous studies will lose accuracy when $\text{Im} \left(\frac{\partial \varepsilon_u}{\partial k_{cj}}\right)\gg \text{Re} \left(\frac{\partial \varepsilon_u}{\partial k_{cj}}\right)$.

\section{Effect of parameter gradients on parametric instability\label{sec_4}}

Under the assumption of weak-inhomogeneity previously adopted, all parameter gradients emerged in Eq. (\ref{Eq_Mbj}) and Eq. (\ref{Eq_V1aj}) , such as  $a_m^\prime$, $b_n^\prime$, etc., were neglected. However, as the inhomogeneity becomes more pronounced, these gradients inevitably influence the results, and will be examined in this section. Unlike wave number mismatch, which exerts its effect primarily through the amplification factor, parameter gradients impact PI by directly altering the local spatial growth rate. Therefore, they must be considered for scenarios where local spatial growth rates are crucial. A typical scenario is the non-resonant type PI triggered within a narrow growth interval $[x_a,x_b]$, whereas the distribution of growth rate is diffuse. In such cases, the amplification is more accurately described by specific integration $\int_{x_a}^{x_b} \text{Re}(r_1) dx$ rather than the amplification factor derived from integration over the entire space.

In this section, we will first analytically examine how these gradients modify the local growth rates by evaluating their effects on coefficients $H_p$. We then use numerical examples from the previous section to further investigate how varying gradients influence the local growth rate and to assess the validity of our approach under different levels of inhomogeneity.

\subsection{Analytical study of the effects }

Considering a moderate level of inhomogeneity where these parameter gradients are non-negligible but small, we revisit Eq. (\ref{Eq_Mbj}) to compute the components of $H_p$'s determinant. Transforming $k_1$ terms in $V_{1ajp}$ to $\alpha$ using Eq. (\ref{Eq_KtoEps}), we obtain the first-order derivative terms of $V_{1ajp}$ as 

\begin{equation}
\begin{split}
    \Delta V_{1ajp}=&\sum_{l=0}^{p}\dfrac{C_{l}^{m}}{(p-l)!}\left[\dfrac{1}{2}\dfrac{\partial^{p-l+2}\alpha}{\partial(ik_{1})^{p-l+2}}(ik_{s})^{m-l}ik_1^\prime+\right.\\
    &\left.\dfrac{d}{dx_1}\dfrac{\partial^{p-l}\alpha}{\partial(ik_{1})^{p-l}}C^{m-l}_1(ik_{s})^{m-l-1}+\dfrac{\partial^{p-l}\alpha}{\partial(ik_{1})^{p-l}}C_2^{m-l}(ik_{s})^{m-l-2}ik_{s}^{\prime}\right].
    \end{split}
\end{equation}

The operator $\dfrac{d}{dx_1}\equiv \dfrac{\partial}{\partial x}+\dfrac{\partial}{\partial k_1}k_1^\prime$, where $\dfrac{\partial \alpha}{\partial x}\equiv\sum_{n=0}^M a_n^\prime (ik_1)^n$ and $\dfrac{\partial \varepsilon_1}{\partial x}\equiv\sum_{n=0}^M b_n^\prime (ik_1)^n$ . For $\Delta V_{1acp}$, the expression is similar to the above, except that $\alpha$ is replaced by $\varepsilon_1$.  

Using the operator $F(b_n^\prime,a_m^{*\prime})$ to denote the contributions of $b_n^\prime$ and $a_m^{*\prime}$ in matrices $\boldsymbol{M}_{b1}$ and $\boldsymbol{M}_{a^*1}$, and simplifying the determinant in Eq. (\ref{Eq_HpDet}), the modification of parameter gradients on the coefficients $H_p$ as
\begin{equation}
    \Delta H_p=\sum_{l=0}^{p}(-i)^{p}\dfrac{\dfrac{\partial^{p-l}}{\partial k_{1}^{p-l}}\dfrac{\partial^{l}}{\partial k_{s}^{l}}}{(p-l)!l!}\left[-\dfrac{ik_{1}^{\prime}}{2}\dfrac{\partial^{2}}{\partial k_{1}^{2}}-i{\dfrac{d}{dx_1}}\dfrac{\partial}{\partial k_{s}}-\dfrac{ik_{s}^{\prime}}{2}\dfrac{\partial^{2}}{\partial k_{s}^{2}}+F(a_m^{*\prime},b_n^\prime)\right]\left(\varepsilon_{1}\varepsilon_{u}-\alpha\alpha^{*}\right).
    \label{Eq_dHp}
\end{equation}

Concluding the general form of $F$ is challenging because $b_n^\prime$ and $a_m^{*\prime}$disrupt the matrix's triangular structure. Consequently, its determinant becomes considerably more intricate and cannot be simplified by extracting a common factor like equation (\ref{Eq_HpWInhom}). For simpler cases, such as a zeroth-order coupling  ($\text{M}=0$) and pure first-order coupling ($\text{M}=1$, $a_0=a_0^*=0$), $\mathrm{det}|\boldsymbol{\bar{M}}_{ab}|$ can take a simpler form as $a_0^{*N}$ and $b_0a_1^{*N-1}$, respectively. The resulting operator $F$ in given by%

\begin{equation}
   F(a_{m}^{*\prime},b_{n}^{\prime})=\begin{cases}
i\dfrac{a_{0}^{*\prime}}{a_{0}^{*}}\dfrac{\partial}{\partial k_{s}} & \text{when }M=0\\
-iD_s\dfrac{\partial}{\partial x_s}+i\dfrac{b_{0}^{\prime}}{b_{0}}D_s+i\dfrac{a_{1}^{*\prime}}{a_{1}^{*}}\dfrac{\partial}{\partial k_{s}} & \text{when }M=1\text{ and }a_{0}^{*}=0
\end{cases}
\label{Eq_Fbnaxm}
\end{equation}

For any function $f$ with respect to $k_s$, the operator $D_s$ is defined by $D_sf(k_s)\equiv \dfrac{f(k_s)-f(0)}{k_s}$. Observing Eq. (\ref{Eq_dHp}) and Eq. (\ref{Eq_Fbnaxm}), five parameter gradients are identified: the gradients of the intrinsic wave numbers $k_0^\prime$, $k_1^\prime$ and $k_u^\prime$, where $k_1^\prime$ and $k_u^\prime$ are equivalent representation of $c_n^\prime$ and $b_n^\prime$, respectively, and the gradients of coupling coefficients $a_m^\prime$ and $a_m^{*\prime}$.

\begin{table}[htbp]
    \centering
\caption{Contributions of various parameter gradients to the modification of the spatial growth rate $\Delta r_1^{\text{od1}}$ for zeroth-order coupling ($\text{M}=0$)}
\label{tab:M0}
    \begin{tabular}{|c|c|c|} \hline 
         \textbf{Parameter gradients} &  $\Delta r_1^{\text{od1}}$ \textbf{from} $\Delta H_{0}$&$\Delta r_1^{\text{od1}}$ \textbf{from} $\Delta H_1$\\ \hline 
         $k_{1}^{\prime}$ or $c_{n}^{\prime}$& 
     $-\dfrac{k_{1}^{\prime}}{2}\dfrac{\partial^{2}\varepsilon_{1}/\partial k_{1}^{2}}{\partial\varepsilon_{1}/\partial k_{1}}$&$ir_{1}^{\text{od1}}\dfrac{k_{1}^{\prime}}{2}\left(\dfrac{\partial^{3}\varepsilon_{1}/\partial k_{1}^{3}}{\partial\varepsilon_{1}/\partial k_{1}}+\dfrac{\partial^{2}\varepsilon_{u}/\partial k_{s}^{2}}{\varepsilon_{u}}\right)$\\ \hline 
 $k_0^\prime$& 0&$ir_{1}^{\text{od1}}\dfrac{k_{0}^{\prime}}{2}\dfrac{\partial^{2}\varepsilon_{u}/\partial k_{s}^{2}}{\varepsilon_{u}}$\\ \hline 
 $a_m^{*\prime}$& 0&$-ir_{1}^{\text{od1}}\dfrac{a_{0}^{*\prime}}{a_{0}^{*}}\dfrac{\partial\varepsilon_{u}/\partial k_{s}}{\varepsilon_{u}}$\\ \hline\end{tabular}

\end{table}

The modification of growth rate $\Delta r_1^{\text{od1}}$ is determined by $\Delta H_0$ and $\Delta H_1$ described by Eq. (\ref{Eq_dHp}). Specifically, $\Delta H_0$ alters $r_1^{\text{od1}}$ by $-\Delta H_0/H_1$, while $\Delta H_1$ contributes $\Delta H_1\frac{H_0}{H_1^2}$. For zeroth order coupling, the growth rate modification is shown in Table \ref{tab:M0}. Most terms are proportional to $r_1^{\text{od1}}$ and appear with the PI coupling. An notable exception is the term driven by $k_1^\prime$ (first row, first column of Table \ref{tab:M0}), representing the linear change of wave amplitude due to the gradient of intrinsic wave number, thus independent of the PI coupling. This term also appears in a single wave equation under inhomogeneity and is the solution of energy flux conservation $(v_{g1}\Phi_1)^\prime =0$.

\begin{table}[h]
    \centering
\caption{Contributions of various parameter gradients to the modification of the spatial growth rate $\Delta r_1^{\text{od1}}$ for purely first-order coupling ($\text{M}=1$ and $a_0=a_0^*=0$.)}
\label{tab:M1}
    \begin{tabular}{|c|c|c|} \hline 
         \textbf{Parameter gradients} &  $\Delta r_1^{\text{od1}}$ \textbf{from} $\Delta H_{0}$&$\Delta r_1^{\text{od1}}$ \textbf{from} $\Delta H_1$\\ \hline 
         $k_{1}^{\prime}$ or $c_{n}^{\prime}$& 
     $-\dfrac{k_{1}^{\prime}}{2}\dfrac{\partial^{2}\varepsilon_{1}/\partial k_{1}^{2}}{\partial\varepsilon_{1}/\partial k_{1}}-ir_{1}^{\text{od1}}\dfrac{k_{1}^{\prime}}{k_{1}k_{s}}$&$ir_{1}^{\text{od1}}\dfrac{k_{1}^{\prime}}{2}\left(\dfrac{\partial^{3}\varepsilon_{1}/\partial k_{1}^{3}}{\partial\varepsilon_{1}/\partial k_{1}}+\dfrac{\partial^{2}\varepsilon_{u}/\partial k_{s}^{2}}{\varepsilon_{u}}\right)$\\ \hline 
 $k_{u}^{\prime}$ or $b_{n}^{\prime}$& $ir_{1}^{\text{od1}}\dfrac{b_{0}^{\prime}}{b_{0}k_{s}}$&$ir_{1}^{\text{od1}}\left(\dfrac{\partial(D_{s}\varepsilon_{u})}{\partial x}\dfrac{1}{\varepsilon_{u}}-\dfrac{b_{0}^{\prime}}{b_{0}}\dfrac{D_{s}\varepsilon_{u}}{\varepsilon_{u}}\right)$\\ \hline 
 $k_0^\prime$& 0&$ir_{1}^{\text{od1}}\dfrac{k_{0}^{\prime}}{2}\dfrac{\partial^{2}\varepsilon_{u}/\partial k_{s}^{2}}{\varepsilon_{u}}$\\ \hline 
 $a_{m}^{\prime}$& $-ir_{1}^{\text{od1}}\dfrac{a_{1}^{\prime}}{a_{1}k_{s}}$&0\\ \hline 
 $a_m^{*\prime}$& $ir_{1}^{\text{od1}}\dfrac{a_{1}^{*\prime}}{a_{1}^{*}k_{s}}$&$-ir_{1}^{\text{od1}}\dfrac{a_{1}^{*\prime}}{a_{1}^{*}}\dfrac{\partial\varepsilon_{u}/\partial k_{s}}{\varepsilon_{u}}$\\ \hline\end{tabular}

\end{table}

For purely first order coupling, the results are shown in Table \ref{tab:M1}. A key difference from Table \ref{tab:M1} is that all five parameter gradients influence the growth rate, whereas $b_n^\prime$ and $a_m^\prime$ do not contribute in the zeroth-order coupling case since Eq. (\ref{Eq_CoupWave1}) is not differentiated when deriving the combined determinant. This suggest that the parameter gradients become more important as the highest differential order $M$ increases.
Aside from the linear term $-\frac{k_1^\prime}{2} \frac{\partial^2\varepsilon_1/\partial k_1^2 }{\partial\varepsilon_1/\partial k_1}$, we can summarize the other terms in Table \ref{tab:M0} and \ref{tab:M1} as $ir_1^{\text{od1}} \frac{P^\prime}{PK_s}$, where $P$ represents the parameter with gradient of interest. Therefore, the term $\frac{P^\prime}{PK_s}$, representing a ratio of adjustment to the growth rate, can serve as an indicator of inhomogeneity.

\subsection{Numerical examples}

We further investigate the effect of parameter gradients on the spatial growth rate in a practical scenario by varying the degree of inhomogeneity and different parameter gradients.
Using the parameters of the non-resonant PI scenario of Sec. \ref{sec_3}, we focus on the local spatial growth rate at the quasi-matching point, where $K_0=13$, $K_1=-9$, and $K_u=4+2i$. We consider the coupling scenarios from the previous analysis: for zeroth order coupling, $a_0=a_0^*=0.3$, and for pure first-order coupling, $ia_1K_1=ia_1^*K_s=0.3$.

To understand how gradients affect the growth rate, we compare the growth rates derived by our approach with and without parameter gradients. We find $\delta^n r_1$ roots of the iterative characteristic polynomials described by Eq. (\ref{Eq_iterate}) and obtain the growth rate $\text{Re}(R_1)$ when $\delta^n r_1$ series converges. For validation, we numerically solve the coupling equations near the quasi-matching point and measure the changing rate of $|\phi_1|$ at the point, representing the growth rate $\text{Re}(R_1)$ of interest. To ensure numerical accuracy, two measures are implemented:\\
First, the endpoint parameters are kept constant by adjusting the solving domain as the inhomogeneity varies. For example, when $K_1^\prime$ increases from 0.001 to 0.002, the domain length is halved to maintain constant $K_1$ at the endpoints. \\
Second, numerical results are filtered to ensure $\phi_1$ accurately describes the amplitude change of the intrinsic mode $k=K_1$. This is crucial as stronger inhomogeneity can excite other intrinsic modes such as $k=-K_1$, even with boundary conditions unrelated to $-K_1$.


\begin{figure}[H]
    \centering
    \includegraphics[width=0.9\linewidth]{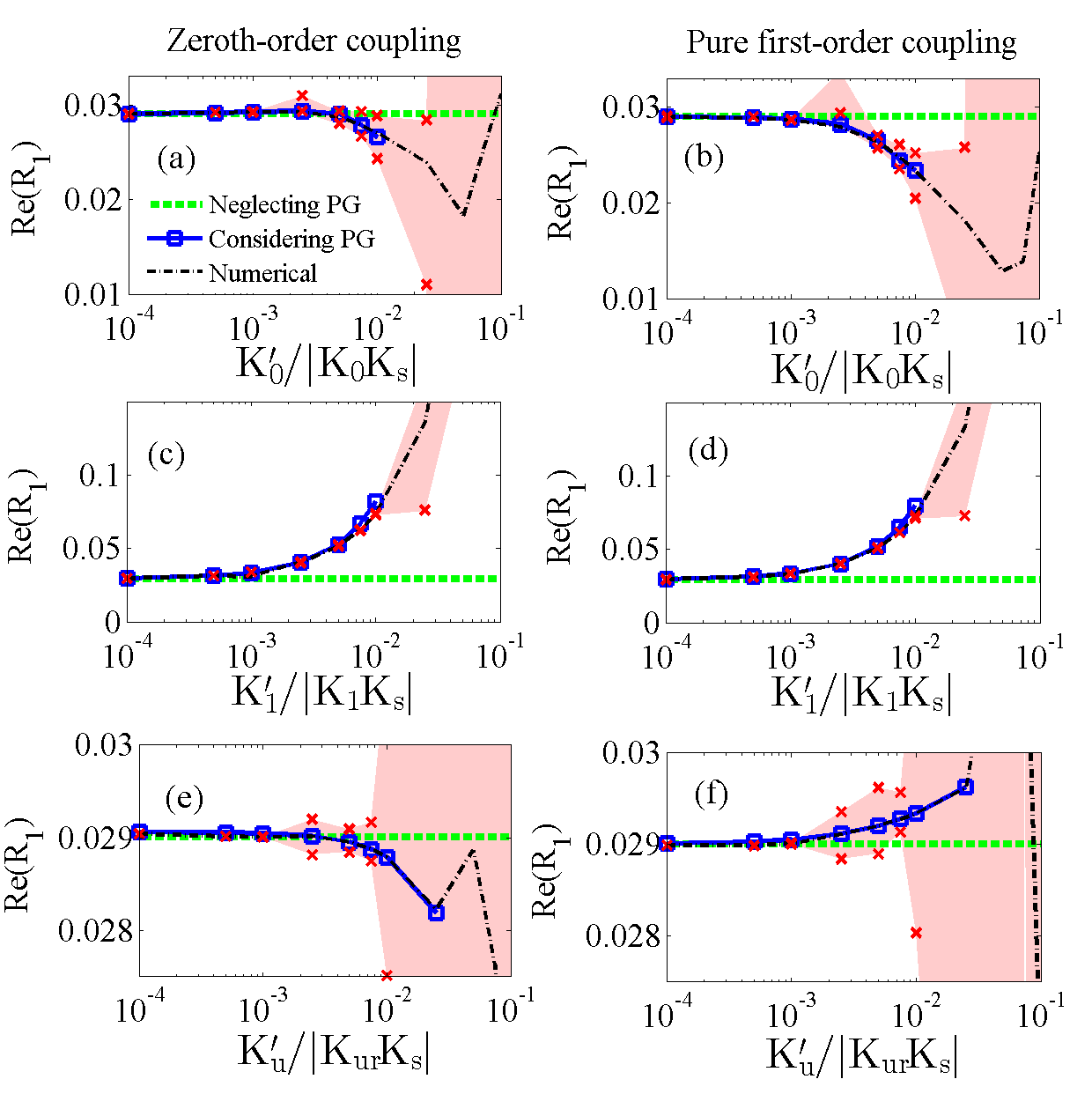}
    \caption{}
\end{figure}
\begin{figure}[H]\ContinuedFloat
    \centering
    \includegraphics[width=0.9\linewidth]{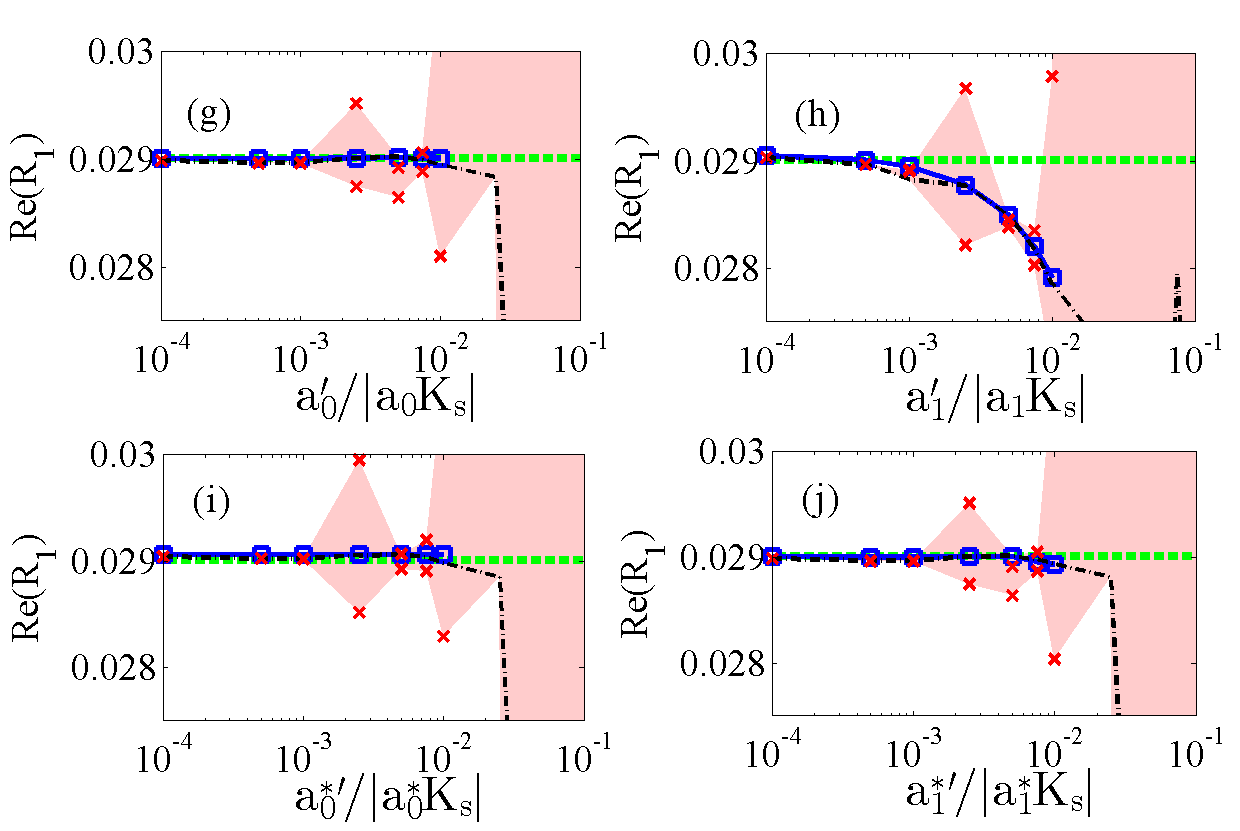}
    \caption[]{\label{fig_PG}Spatial growth rate $\mathrm{Re(R_1)}$ at the quasi-matching point calculated using our approach without parameter gradients (green dashed line), our approach with parameter gradients (blue square line) and numerical solutions (black dot-dashed line) versus the degree of inhomogeneity. Each row corresponds to different parameter gradient, with other gradients set to zero. Red area outlined by crosses represents the sensitivity of the numerical solution to its boundary condition. The left column shows results for the zeroth-order coupling scenario ($\text{M}=0$), while the right column presents results of pure first-order coupling scenario ($\text{M}=1$, $a_0=a_0^*=0$). Parameter gradient is abbreviated as PG in the legend.}
\end{figure}
In Fig. \ref{fig_PG}, the growth rate without parameter gradients (green dashed line) remains constant at 0.029, regardless of the inhomogeneity. The difference between this constant and the growth rates with parameter (blue line and squares) indicates the effect of parameter gradients on growth rate.
Across all panels of Fig. \ref{fig_PG}, when the inhomogeneity is very weak ($10^{-4}\sim10^{-3}$), the impact of the parameter gradient is minimal. However, as the inhomogeneity becomes stronger ($10^{-3} \sim 10^{-2}$), the blue and green lines diverge, showing a significant effect of parameter gradients. Numerical results (black dot-dashed line) in this range support the result of model considering parameter gradients. For inhomogeneity larger than $10^{-2}$, the $\delta^n r_1$ series described by Eq. (\ref{Eq_iterater1}) gradually diverge, and no consistent $R_1$ solution is observed, resulting in almost no blue data in this range. 

To ensure the reliability of numerical growth rate, we also evaluated its stability against boundary condition variations. In Fig. \ref{fig_PG}, additional numerical growth rates (red crosses) under boundary conditions 0.9 or 1.1 times of the original one are shown. The red area outlined by these growth rates indicates the sensitivity to the boundary condition. Across all panels, the numerical growth rate remains nearly unaffected by the boundary condition for inhomogeneity below $10^{-2}$. However, for inhomogeneity above this threshold, the red area expand significantly, showing that the local growth rate becomes highly dependent on the boundary condition loses stability, making it impossible to describe the wave '1' behavior with merely a growth rate.

Fig. \ref{fig_PG} details align with the previous analytical analysis. The effect of parameter gradients for pure first-order coupling (panels of right column) is generally stronger than that for zeroth-order coupling (panels of left column), reflecting the larger number of adjusting terms found in Table \ref{tab:M1}. The most significant effect is driven by the linear term $-K_1^\prime/2K_1$, changing the growth rate from 0.029 to 0.080 when the inhomogeneity increases to $10^{-2}$. Other parameter gradient  adjustments are much smaller, consistent with the common analytical form $ir_1^{\text{od1}}\frac{P^\prime}{Pk}$. The largest change, aside from that induced by $-K_1^\prime/2K_1$, is observed in Fig. \ref{fig_PG}(b), from 0.029 to 0.023, about $21\%$ of the unaffected growth rate.

\section{Conclusions and Discussions\label{sec_5}}

The extensively studied three wave interactions, a form of parametric instability (PI), involve a pump wave splitting into two child waves named as wave '1' and wave 'u' in this study. This work presents an analytical approach to solve the governing coupling wave equations based on a WKBJ method to comprehend the PI process in inhomogeneous media. The approach involves eliminating the perturbation $\phi_u$ of mode 'u' and expanding the perturbation $\phi_1$ of wave '1' using the WKBJ approximation. The wave coupling equations then transform into the amplitude equation $\sum_p H_p\Phi_1^{(p)}=0$ for the amplitude of child wave '1'. The solution, represented by the growth rate $R_1$, is constructed by series $r_1+\delta r_1+\delta^2 r_1...$, where $\delta^n r_1$ is the minimal root of $n$th iterative characteristic polynomials described by Eq. (\ref{Eq_iterate}). Under very weak inhomogeneity, the amplitude equation is equivalent to the Taylor expansion of parametric dispersion relation, and $r_1$ suffices to depict the growth rate. As inhomogeneity strengthens, parameter gradients and higher-order $\delta^n r_1$ components must be considered.

The effect of wave number mismatch on PI is investigated based on solutions of the amplitude equation under weak inhomogeneity. The growth rates of non-resonant and resonant type PI are described by the minimal root of the characteristic polynomial truncated at the first and second order, denoted as $\text{Re}(r_1^{\text{od1}})$ and $\text{Re}(r_1^{\text{od2}})$, respectively. By reinterpreting the resulting key quantities, physical aspects of various PI types of are unified:\\
(i) The intrinsic wave number of the quasi-mode should be a complex root of its dispersion equation $\varepsilon_u(k)=0$, similar to that the real root of $\varepsilon_1(k)=0$ is the intrinsic wave number of the resonant wave '1'.\\
(ii) The wave number mismatch of both types PI is the difference between the intrinsic wave numbers of the three waves. For each root $k_j$ of dispersion equation $\varepsilon_u(k)=0$, the wave number mismatch is
\begin{equation}
    \kappa_j=k_j-k_0-k_1.
\end{equation}
(with signs of $k_0$ and $k_1$ following frequency matching rules.) The matching point $z_j$ is the root of the complex function $\kappa_j(z)=0$. For the non-resonant type PI, $z_j$ is not on the real axis, but a quasi-matching point $x_j$ satisfying $\mathrm{Re}[\kappa_j(x_j)]=0$ can be found, where the spatial growth rate maximum.\\
(iii) The amplification factor depends on the matching points over the space. Combining Eqs. (\ref{Eq_AinhomReNew}), (\ref{Eq_AinhomNrNew1}) and (\ref{Eq_AinhomItBw}), the contribution the $j$th matching point the amplification factor, including resonant type ($\text{Im}(k_{j})=0$), non-resonant type ($\text{Im}(k_j)\neq 0$), forward scattering ($v_{g1}v_{gu}>0$) and backward scattering ($v_{g1}v_{gu}<0$),  is unified as
\begin{equation}
    A_{\text{inhom}}|_j=\begin{cases}
\text{sign}(\kappa_{rj}^{\prime})\text{Re}\left(\dfrac{-\pi\Lambda_{j}}{\kappa_{j}^{\prime}}\right) & \text{Im}(k_{j})\ge0,\text{forward}\\
-2\text{Im}(k_{j})\dfrac{\sqrt{\Lambda_{j}}}{|\kappa_{j}^{\prime}|} & \text{Im}(k_{j})\in[-\frac{\pi}{2}\sqrt{\Lambda_{j}},0),\text{backward}\\
\text{sign}(\kappa_{rj}^{\prime})\text{Re}\left(\dfrac{\pi\Lambda_{j}}{\kappa_{j}^{\prime}}\right) & \text{Im}(k_{j})<-\frac{\pi}{2}\sqrt{\Lambda_{j}},\text{backward}
\end{cases}
\end{equation}
where $\Lambda_j\equiv -\dfrac{\gamma_0^2}{v_{g1}v_{guj}}$.

The effect of parameter gradients on the local spatial growth rate is also investigated using our approach. Analytical analysis reveal that the wave number gradient of child wave '1' influences its amplitude independent of PI, resulting from the linear response of wave to the inhomogeneity. Other gradient-induced effects are proportional to the PI spatial growth rate, where a gradient $P^\prime$ can alter the growth rates in rates of $P^\prime/|Pk_s|$. Our numerical examples validate our approach under varying degree of inhomogeneity indicated by $P^\prime/|Pk_s|$. When the degree of inhomogeneity is below $10^{-3}$, our model without parameter gradients is valid. Including parameter gradients extends this validity to an inhomogeneity of about $10^{-2}$.


Although this work focused on wave interactions in 1D media, the results may be applicable in 2D and 3D media, where the wave equations are partial differential equations. One approach is to extract a linear ordinary equation with respect to the inhomogeneous dimension from the governing coupling equations, assuming the symmetry in the other dimensions. Another approach is to first obtain the amplitude equation using the WKBJ approximation and then solve it with practical boundary conditions. Future studies will explore application of these approach in specific wave-plasma scenarios.

\begin{acknowledgments}
This work is supported by the National Natural Science Foundation of China (Grant Nos. 12175274, 12175275 and 12375231), the Youth Innovation Promotion Association Chinese Academy of Sciences (Y2021114) and Collaborative Innovation Program of Hefei Science Center, CAS, No. 2022HSC-CIP023. 
\end{acknowledgments}

\bibliography{MyRefs.bib}
\bibliographystyle{unsrt}

\end{document}